\documentclass[preprint]{vgtc}             

\ifpdf
  \pdfoutput=1\relax                   
  \usepackage{graphicx}                
  \usepackage{hyperref}
  \DeclareGraphicsExtensions{.pdf,.png,.jpg,.jpeg} 
\else
  \usepackage{graphicx}                
  \DeclareGraphicsExtensions{.eps}     
\fi%

\graphicspath{{figures/}{pictures/}{images/}{./}} 
\usepackage{pbox}
\usepackage{microtype}                 
\PassOptionsToPackage{warn}{textcomp}  
\usepackage{textcomp}                  
\usepackage{mathptmx}                  
\usepackage{times}                     
\usepackage{cite}                      
\usepackage{tabu}                      
\usepackage{booktabs}                  
\usepackage{multirow}
\usepackage[utf8]{inputenc}

\onlineid{0}

\vgtccategory{Research}

\vgtcinsertpkg

\usepackage{subcaption}
\usepackage{todonotes}
\usepackage{amssymb}
\usepackage{tabularx}
\usepackage{makecell}
\usepackage{graphbox}
\usepackage{algorithmic}
\usepackage{algorithm}

\newcommand{\scenery}{\emph{scenery} \cite{scenery}}
\newcommand{\insitu}{\textit{in situ}}
\newcommand{\sensitivity}{$\gamma$}
\newcommand{\criteria}{$\tau$}
\newcommand{\superseg}[2]{\ensuremath{\mathbb{S}^{#1}_{#2}}}
\newcommand{\supseglist}[1]{\ensuremath{\mathbb{L}_{#1}}}
\newcommand{\front}[2]{\ensuremath{f(\mathbb{S}^{#1}_{#2})}}
\newcommand{\back}[2]{\ensuremath{b(\mathbb{S}^{#1}_{#2})}}
\newcommand{\vieworig}[0]{\ensuremath{\mathrm{V}_\mathrm{O}}}
\newcommand{\viewnew}[0]{\ensuremath{\mathrm{V}_\mathrm{N}}}
\newcommand{\numSupsegs}[0]{\ensuremath{\mathbb{N_{S}}}}
\newcommand{\numLists}[0]{\ensuremath{\mathbb{N_{L}}}}
\newcommand{\downimage}{\ensuremath{D_I}}
\newcommand{\downray}{\ensuremath{D_R}}
\newcommand{\by}{$\times$}
\newcommand{\fhdres}{$1920\times1080$}
\newcommand{\shdres}{$1280\times720$}

\ieeedoi{10.1109/PacificVis56936.2023.00014}

\title{Efficient Raycasting of Volumetric Depth Images for Remote Visualization of Large Volumes at High Frame Rates}

\author{Aryaman Gupta\thanks{e-mail: aryaman.gupta@tu-dresden.de}\\ %
        \parbox{1.8in}{\scriptsize \centering
        Technische Universit\"at Dresden \\
        Center for Systems Biology Dresden \\
        MPI-CBG, Dresden}%
\and Ulrik G\"unther\\ %
     \parbox{1.8in}{\scriptsize \centering
        CASUS, G\"orlitz \\
        Center for Systems Biology Dresden \\
        MPI-CBG, Dresden} %
\and Pietro Incardona\\ %
     \parbox{1.8in}{\scriptsize \centering
        Technische Universit\"at Dresden \\
        Center for Systems Biology Dresden \\
        MPI-CBG, Dresden}
\vspace{1.2em}
\and Guido Reina\\ %
     \parbox{1.3in}{\scriptsize \centering
        Visualization Research Center, University of Stuttgart}
\and Steffen Frey\\ %
     \parbox{1.2in}{\scriptsize \centering
       University of Groningen \\}
\and Stefan Gumhold\\ %
     \parbox{1.6in}{\scriptsize \centering
        Technische Universit\"at Dresden \\}
\and Ivo F. Sbalzarini\thanks{e-mail: ivo.sbalzarini@tu-dresden.de}\\ %
     \parbox{1.8in}{\scriptsize \centering
        Technische Universit\"at Dresden \\
        Center for Systems Biology Dresden \\
        MPI-CBG, Dresden}}

\teaser{
    \centering
    \includegraphics[width=\textwidth]{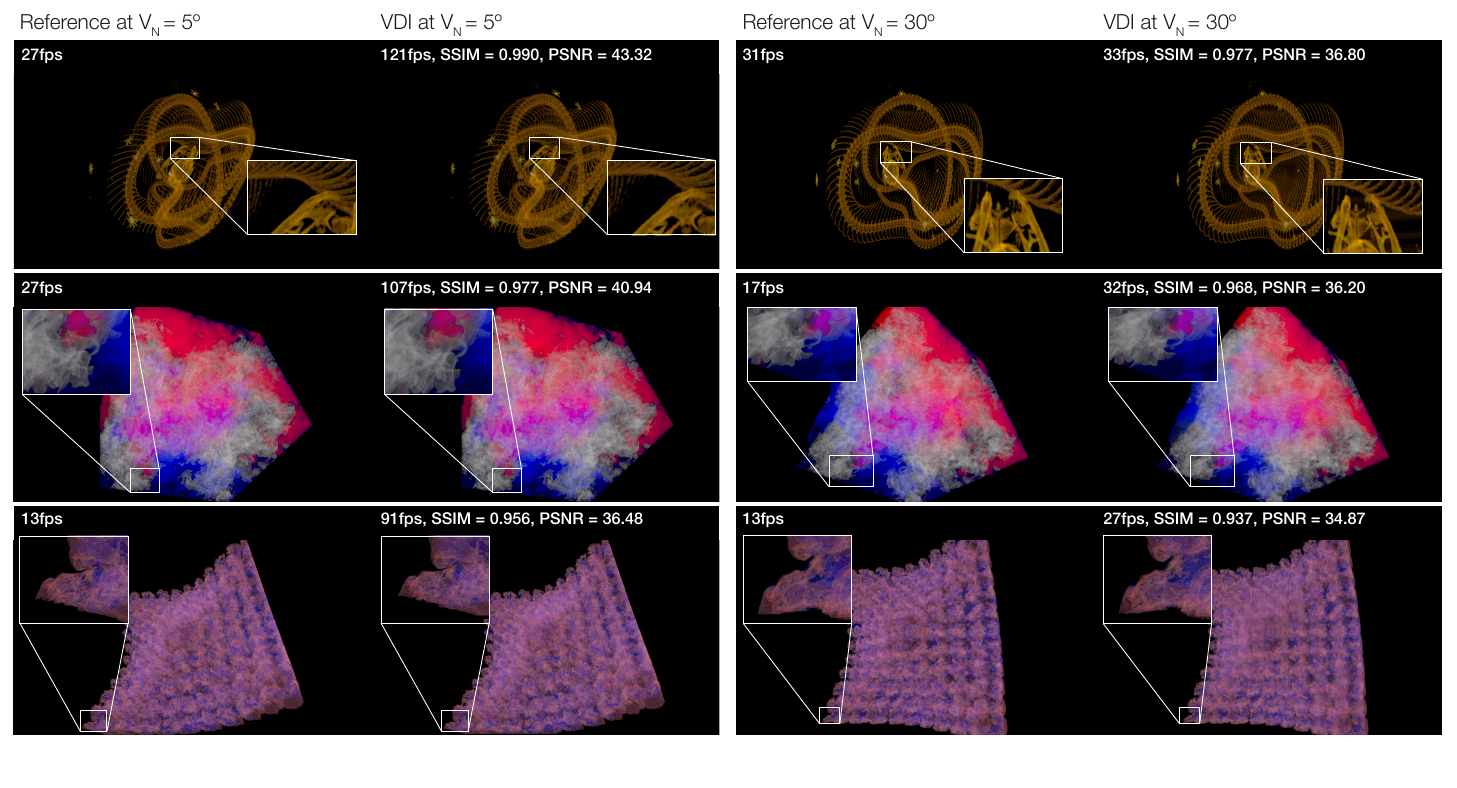}
    \caption{Visual comparison of VDI rendering quality with direct volume rendering (DVR) for the Kingsnake (top), Rayleigh-Taylor (middle), and Richtmyer-Meshkov (bottom) datasets. All frame rates are reported without empty-space skipping for comparability. Image quality metrics are computed w.r.t.~the DVR image at the same angle.}
    \label{fig:image_comparisons}
}

\abstract{
We present an efficient raycasting algorithm for rendering Volumetric Depth Images (VDIs), and we show how it can be used in a remote visualization setting with VDIs generated and streamed from a remote server. VDIs are compact view-dependent volume representations that enable interactive visualization of large volumes at high frame rates by decoupling viewpoint changes from expensive rendering calculations. However, current rendering approaches for VDIs struggle with achieving interactive frame rates at high image resolutions.
Here, we exploit the properties of perspective projection to simplify intersections of rays with the view-dependent frustums in a VDI and leverage spatial smoothness in the volume data to minimize memory accesses. Benchmarks show that responsive frame rates can be achieved close to the viewpoint of generation for HD display resolutions, providing high-fidelity approximate renderings of Gigabyte-sized volumes. We also propose a method to subsample the VDI for preview rendering, maintaining high frame rates even for large viewpoint deviations. We provide our implementation as an extension of an established open-source visualization library.

} 

\CCScatlist{
  \CCScatTwelve{Human-centered computing}{Visu\-al\-iza\-tion}{Visualization theory, concepts and paradigms}{}
  \CCScatTwelve{Human-centered computing}{Visu\-al\-iza\-tion}{Visualization techniques}{}
}

\begin{document}



\maketitle

\section{Introduction} 

Interactive direct volume rendering is commonly used in the exploration and analysis of three-dimensional volume data. Rendering at high, consistent frame rates is crucial for enabling interactive viewpoint changes and zooming, which are important for gaining depth perception and spatial understanding.
As scientific simulations and experimental devices generate larger data, however, it is increasingly challenging to achieve high, consistent volume rendering frame rates. In remote visualization applications, such as live \insitu~visualization of numerical simulations, fluent user interaction is potentially also hindered by network latency to the server. 

View-dependent, piecewise constant representations of volumetric data, also known as Volumetric Depth Images (VDIs) \cite{Frey}, provide a potential solution by decoupling expensive rendering from interactive viewpoint changes. These representations decompose the volume rendering integral into segments that store composited color and opacity. Rendering such a representation involves compositing these segments, which is less expensive than performing the full integration \cite{overview} and produces high-fidelity approximations for camera viewpoints near the viewpoint from which the representation was generated \cite{Frey, novelview}. Additionally, VDIs are more compact than the original volume data and can be generated and streamed efficiently \cite{Frey, spacetime}.
This provides an attractive potential solution for interactive remote rendering. 
However, the large number of segments and their shape (pyramidal frustums when using perspective projection) makes VDIs challenging to render efficiently. Existing VDI rendering methods are thus unable to maintain interactive frame rates for high-definition (HD) displays.

Here, we present an efficient raycasting-based VDI rendering method  designed to scale to large volumes and high-resolution (full HD) displays. At the core of our raycasting algorithm is a simplified way of intersecting rays with segments by computing them in Normalized Device Coordinate (NDC) space, as well as minimizing memory accesses by exploiting spatial homogeneity in the data. We additionally show how empty regions can be skipped in VDI raycasting. In comparison to the previous state-of-the-art object-order rendering algorithm for VDIs \cite{Frey}, we report frame rates that are an order of magnitude higher near the viewpoint of generation, while maintaining the same rendering quality with respect to ground-truth volume rendering. In comparison to a previously proposed raycasting approach \cite{novelview}, our method further reduces calculations and memory accesses.

Generating a VDI requires a dataset and transfer-function dependent sensitivity parameter to control the partitioning of rays into segments.
In previous works, this parameter needed to be tuned manually, further hampering interactivity.
We here instead propose a technique to automatically optimize the VDI generation parameter for given constraints. Importantly, this enables the parameter to be individually tuned for each ray, generating effectively content-adaptive VDIs that provide better-quality visualizations.

We particularly target a use case in which VDIs are generated at a remote server from the user's most recent viewpoint and streamed to a visualization client.
A VDI at the client enables local renderings of user interactions until an updated VDI becomes available from the server. 
To account for the fact that the rendering frame rate decreases with increasing deviation in camera viewpoint---due to the anisotropic shape of the VDI---, we propose a technique to adaptively adjust the sampling along a ray so as to maintain a set frame rate.

In summary, we contribute an efficient raycasting method for VDIs that outperforms the current state of the art \cite{Frey, novelview}. We also suggest 
 a downsampling method for preview rendering at set frame rates and
a method to automatically adjust the VDI generation parameter individually for each ray, removing the need for manual tuning.  We benchmark the proposed algorithms on several datasets and provide an open-source implementation as part of the visualization library \scenery.

\begin{figure*}[t!]
    \centering
    \begin{subfigure}[t]{0.49\textwidth}
        \centering
        \includegraphics[height=2in]{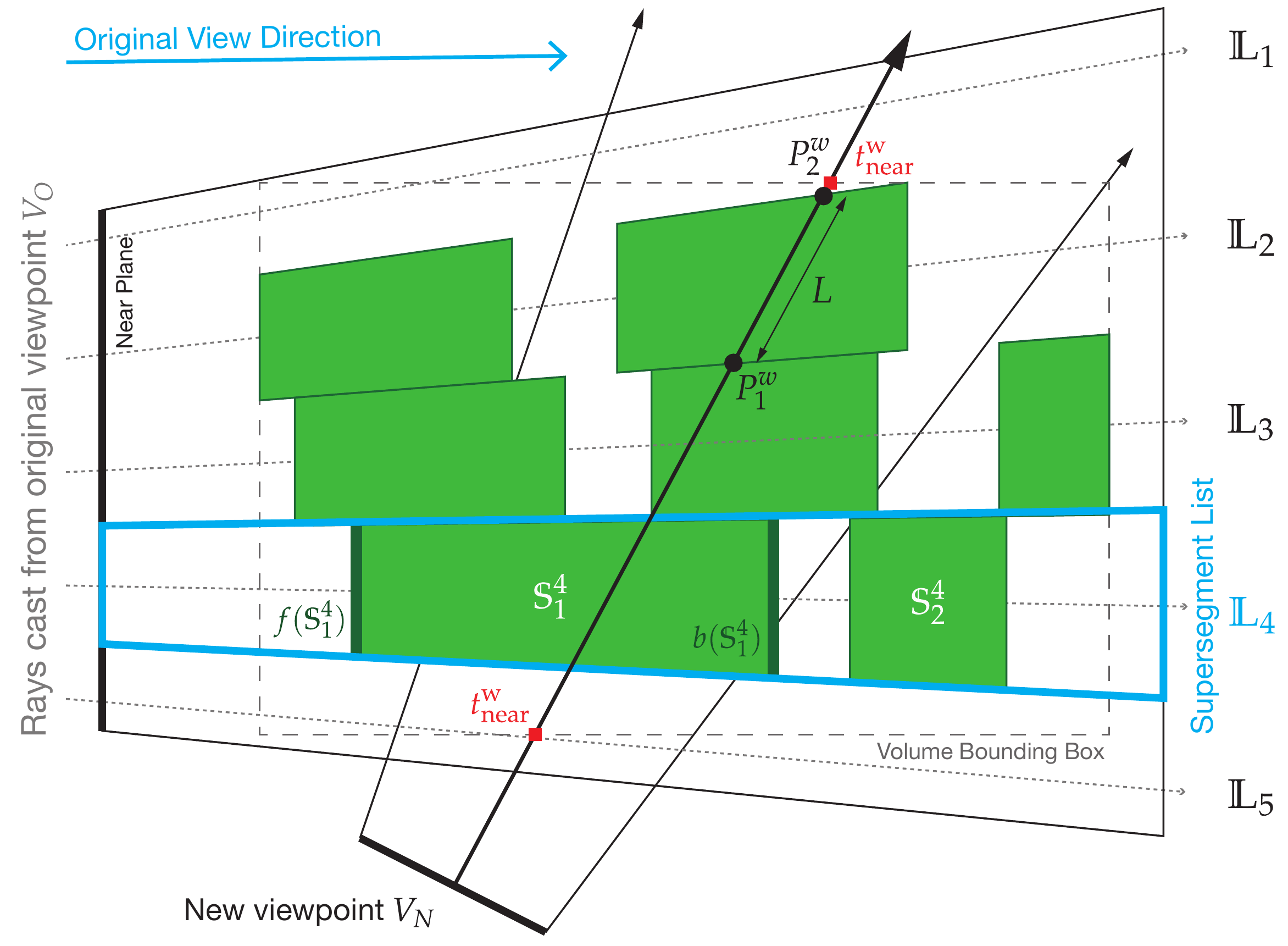}
        \caption{\textbf{World space}. The structure of a Volumetric Depth Image \cite{Frey} and the rendering calculations that are performed in world space. \superseg{}{} and \supseglist{} form pyramidal frustums in world space. The start and end points for ray marching are determined by intersecting the viewport and the volume bounding box. Once \superseg{i}{j} intersection points are determined in NDC space, they are converted back to world space to determine the length of intersection.}
        \label{fig:vdistructure}
    \end{subfigure}%
    \hfill 
    \begin{subfigure}[t]{0.49\textwidth}
        \centering
        \includegraphics[height=1.9in]{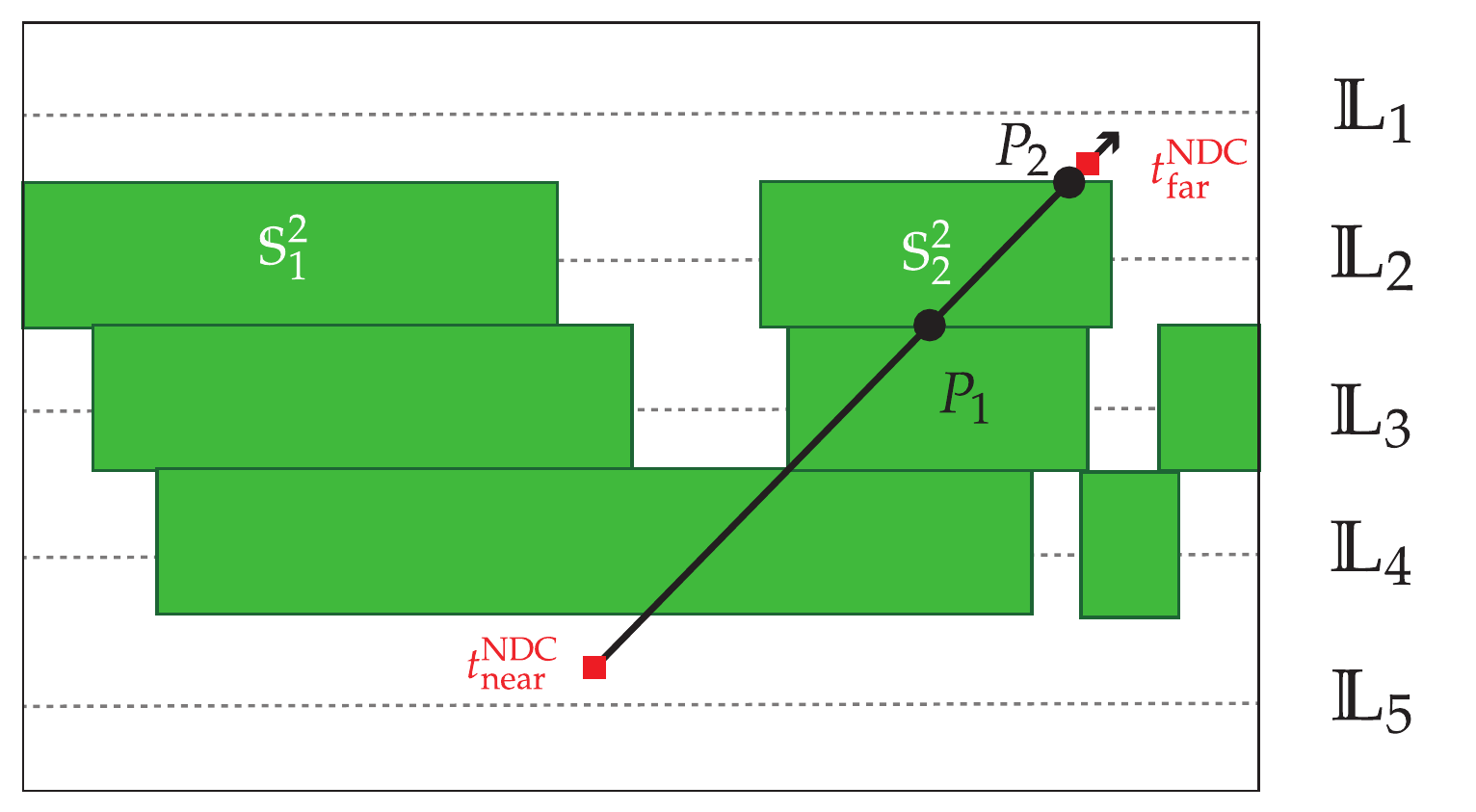}
        \caption{\textbf{NDC space}. Raycasting the VDI in NDC space of {\vieworig} for determining \superseg{}{} intersections. The start and end points of the ray are converted to NDC space, and the ray steps through \supseglist{}, determining \superseg{}{} intersections. The intersection points are converted back to world space to determine intersection length.}
        \label{fig:clipspace}
    \end{subfigure}
    \caption{The structure of a Volumetric Depth Image (VDI) and the coordinate transformations involved in VDI rendering.}
    \label{fig:raycasting}
\end{figure*}

\section{Related Work and Background}\label{sec:related}

Recent work in remote rendering has leveraged hardware-accelerated video encoding \cite{cuttingcord, remotehwenc}, client-side reprojections \cite{remoteprojectns, remoteprojectns2}, and motion prediction \cite{motionprediction} to achieve high frame rates and low interaction latency. However, for direct volume rendering of large data, the primary bottleneck remains the rendering time itself. We review the literature in explorable image representations (\autoref{sec:related-explorable}) and deep-learning novel view synthesis (\autoref{sec:related-deep}), which both provide potential solutions. Then, we provide some background about the specific explorable image representation we use here, the VDI \cite{Frey} (\autoref{sec:related-vdi}).




\subsection{Explorable Image Representations}\label{sec:related-explorable}
Several explorable image representations have been proposed in the literature, often with the goal of decoupling rendering from interaction in remote-rendering applications. Shade et al. \cite{layered} introduced the view-dependent Layered Depth Image (LDI), storing multiple pixels along each line of sight. This allows for deferred rendering, but is limited to surface and geometry data. Stone et al. \cite{omni} rendered and streamed omni-directional stereoscopic images of molecular dynamics simulations from remote compute clusters, using local reprojections at frame rates suitable for Virtual Reality (VR). However, omni-directional stereoscopic images require warping to prevent distortions \cite{warping}, which depends on depth information and therefore cannot be applied to volume data. For reprojecting volume data, Zellmann et al. \cite{zellmann, zellmann2} transmitted a single depth layer along with the color buffer from the rendering server and provided a number of heuristics to generate the depth buffer. While the use of a single depth value per pixel minimizes message sizes, it only yields low-quality reprojections with visible holes where rays do not intersect the depth layer. VDIs were introduced by Frey et al. \cite{Frey} and store a piecewise constant discretization of the volume rendering integral with no gaps or holes. They have been shown to produce higher-quality renderings \cite{Frey}.

Tikhonova et al. \cite{tikhonova_multiple, tikhonova_coherent} proposed compact view-dependent representations that support interactive transfer function changes.
Recently, Rapp et al. \cite{Rapp} modeled scalar densities along each ray in the Fourier domain, generating more compact representations of the volume while still supporting interactive transfer function changes. The focus of these works, however, is on interactive transfer function changes. This differs from the works mentioned in the first paragraph, which all consider a given transfer function, but focus on rendering speed from novel viewpoints. While Rapp et al. \cite{Rapp} do support viewpoint changes, they require slow bilinear interpolation of Lagrange multipliers.

In another fundamentally different approach, Ahrens et al. \cite{cinema} proposed {\it Cinema} for \textit{post hoc} explorative visualization of numerical simulations from a database of images generated \insitu~using different visualization parameters and camera viewpoints. All parameters and viewpoints, however, must be specified in advance, and the database becomes large if many viewpoints are considered. 

\subsection{Deep Learning for Novel View Synthesis}\label{sec:related-deep}
More generally, explorable image representations can be seen as approaches to novel view synthesis, i.e., to the problem of using ``images'' of a scene to generate an image from a new viewpoint. In recent years, this problem has also been addressed using deep learning techniques.
Mildenhall et al. \cite{nerf}, for example, proposed the NeRF (Neural Radiation Fields) representation. Their neural network encodes a continuous volume using weights approximating pre-classified  RGB$\alpha$ values at any point in space.
More recent works explored implicit neural representations of large  volume data, achieving high compression ratios \cite{nnvolume, nnvolume2, weiss}. Unlike NeRF, these methods do not approximate classified values (i.e., with the transfer function applied), but directly predict the data at the query point.

Importantly, implicit neural representations can be rendered from new viewpoints by raycasting. Collecting samples by neural inference, however, is slow \cite{nerf, nnvolume}. Higher frame rates are achieved by efficiently distributing samples along the ray \cite{ngp}, varying the step size using an acceleration data structure \cite{nnvolumefast}, or by sampling a discretized grid \cite{fastnerf} or octree \cite{plenoctrees}. Large, dense regions in volumes, however, still require many samples to be taken, limiting the frame rate. The present VDI approach is complementary, as it could, in fact, be used to cache a neural representation for efficient rendering by VDI raycasting at high frame rates.

\subsection{Volumetric Depth Image and Rendering}\label{sec:related-vdi}
A VDI is a compressed, view-dependent piecewise constant explorable representation of volumetric data. We recall it here using the original notation by Frey et al. \cite{Frey}. \autoref{fig:vdistructure} illustrates the structure of a VDI. Each ray cast into the volume from the original viewpoint (\vieworig) decomposes the rendering integral into so-called \emph{supersegments}. 
Each ray $i$ generates a list \supseglist{i} of supersegments \superseg{i}{j} that store the distances between the near-plane and their front and back faces, \front{i}{j} and \back{i}{j}, respectively.
Each \superseg{i}{j} contains classified accumulated color and opacity between \front{i}{j} and \back{i}{j}. Fully transparent regions in the volume are not included in supersegments.

Given perspective projection during VDI generation, \supseglist{} subdivide the space spanned by the perspective view frustum.
As such, all \supseglist{i} and \superseg{i}{j} are irregular pyramidal frustums. During VDI generation, the number of supersegments in each list, \numSupsegs, is limited to a pre-set maximum \cite{Frey, novelview, brady}. This, together with the viewport resolution, i.e., the number of lists \numLists, determines the size of the VDI. 
Typically, the number of lists is much larger than the number of supersegments per list, i.e., \numLists $\gg$ \numSupsegs.

Different criteria have been proposed to determine supersegment lengths. Brady et al. \cite{brady} generate equal sized supersegments along each ray. This, however, composites supersegments over potentially highly heterogeneous samples, hampering the quality of rendering from a new viewpoint \viewnew. Lochmann et al. \cite{novelview} therefore divide the accumulated opacity along a ray equally among the supersegments. This, however, does not account for potentially varying color values within supersegments. Frey et al. \cite{Frey} generated accurate VDIs using homogeneity as the criterion for supersegment lengths. Samples along the ray are accumulated into a supersegment until they differ from the supersegment by more than a user-defined threshold \sensitivity, in which case a new supersegment is started. Here, we extend this method by automatically determining \sensitivity~separately for each ray, eliminating manual tuning and further improving quality.

Rendering a VDI requires integrating over the \superseg{i}{j} instead of the original data voxels.
Several methods have been proposed for this. Brady et al. \cite{brady} rely on equal-sized supersegments in each list for their alpha-blending rendering. This, however, cannot generalize to supersegments of arbitrary lengths. 
Frey et al. \cite{Frey} thus proposed an object-space approach that creates a frustum geometry for each supersegment \superseg{i}{j}. 
The supersegment lists \supseglist{i} are then sorted from the new viewpoint and rendered using alpha blending.
The opacity contribution from \superseg{i}{j} is based on the intersection length with a ray from the new viewpoint. 
This approach, however, requires creating six triangles for each \superseg{i}{j}, becoming prohibitive for large VDIs.


Ray-based techniques do not explicitly create any geometry and therefore scale better to large VDIs. 
They also allow for early ray termination and can better leverage the anisotropy of the VDI, as rays can march quickly along the lists.
In the raycasting method by Lochmann et al. \cite{novelview}, a ray from \viewnew~is projected onto \vieworig, rasterized, and traversed using a DDA (digital differential analyzer) to determine the intersected lists. 
It then computes intersections with the pyramidal frustums of the  supersegments. 

The present raycasting method instead computes \superseg{i}{j} intersections in the Normalized Device Coordinate (NDC) space of \vieworig, where all \superseg{i}{j} and \supseglist{i} are cuboids (\autoref{fig:clipspace}). This enables the use of voxel stepping \cite{amanatides} to traverse \supseglist{} and simplifies the computation of the intersections. Additionally, we exploit the spatial smoothness across lists to reduce memory accesses, and we propose methods to skip empty regions and to sub-sample rays for preview rendering.

\begin{table}[]
\centering
\begin{tabular}{l|l|l}
\hline
Dataset & Dimensions & Datatype \\ \hline \hline
Engine & $256\times256\times256$ & uint8 \\ 
Kingsnake & $1024\times1024\times795$ & uint8 \\ 
Rayleigh-Taylor \cite{rayleigh} & $1024\times1024\times1024$ & uint16 \\ 
Richtmyer-Meshkov \cite{richtmyer} & $2048\times2048\times1920$ & uint8 \\ \hline
\end{tabular}
\caption{Datasets used to evaluate the presented algorithms.}\label{tab:datasets}
\end{table}

Throughout this manuscript, we rationalize our design decisions in benchmarks using the four volume datasets listed in \autoref{tab:datasets}.
The Engine dataset is a CT scan of two cylinders of an engine block.
The Kingsnake dataset is an X-ray CT scan of an egg of the {\it Lampropeltis getula} snake species. The Rayleigh-Taylor dataset is the density field at a single time step of a computer simulation of the fluid instability of same name \cite{rayleigh}. The Richtmyer-Meshkov dataset is the entropy field at a single simulation time step of the so-named instability \cite{richtmyer}.
Unless otherwise stated, measurements were done on a workstation with an Nvidia GeForce RTX 3090 GPU and an AMD Ryzen Threadripper 3990x 64-core CPU running Ubuntu 20.04. Volume raycasting, both for generating VDIs and for direct volume rendering (DVR), used an emission-absorption illumination model. Due to a buffer size limitation in the runtime system, datasets larger than 2\,GB were distributed across multiple buffers for DVR.

\section{Ray-Adaptive Generation of Supersegments}\label{sec:generation}
In order to generate accurate VDIs, Frey et al. \cite{Frey} proposed a homogeneity criterion \criteria~for supersegment termination:
\begin{equation}\label{eq:termination}
   \tau : \gamma > || C(\mathbb{S})\alpha(\mathbb{S}) - C'\alpha' ||_2,
\end{equation}
where $C'$ and $\alpha'$ are the color and the length-adjusted opacity of the next sample. In words, a sample along the ray is merged into the current \superseg{}{} unless it differs from the pre-multiplied color of \superseg{}{} by more than a user-defined sensitivity parameter \sensitivity, in which case a new \superseg{}{} is started. This criterion generates homogeneous \superseg{}{}, but the sensitivity parameter \sensitivity~is constant across rays and must be carefully tuned manually for a given dataset and transfer function so as to prevent visual artefacts. Too high values of \sensitivity~generate insufficient supersegments to represent the data. Too low values exhaust the supersegment budget \numSupsegs, potentially causing ``smearing'' artefacts as the last \superseg{}{} must contain all remaining data. We address this issue by proposing a method to automatically determine a suitable  \sensitivity~independently for each ray, while guaranteeing a maximum of \numSupsegs~supersegments per ray. 

Leveraging the fact that the number of supersegments produced decreases monotonically with increasing \sensitivity, we perform bisection search between the highest and lowest possible values of \sensitivity~until a value is found that generates \numSupsegs~supersegments, i.e., we optimize the resolution along rays without violating the constraint \numSupsegs~(see Supplement for pseudo-code of the algorithm). Since the distance metric in \autoref{eq:termination} is an $L_2$ distance between pre-multiplied color vectors with 3 elements each, the highest possible \sensitivity~value is $\sqrt{3}$ and the lowest is $0$.
Each iteration of bisection search samples the volume along the ray to determine the number of supersegments generated for the current \sensitivity. Since many iterations may be required to 
determine a \sensitivity~that generates exactly \numSupsegs~supersegments, a tolerance of up to $\delta$ fewer supersegments than \numSupsegs~is permitted, but never more than \numSupsegs~as this would cause ``smearing'' artefacts. We empirically find a $\delta$ of 15\% of \numSupsegs~to provide a good trade-off between performance and quality. To eliminate rays that pass through empty or homogeneous regions, we initialize \sensitivity~to a small positive value, here $10^{-5}$. If the first iteration then generates fewer supersegments than \numSupsegs, the samples along that ray are homogeneous, and that ray can immediately terminate,  freeing computational resources for other rays.
Measured VDI generation times for different datasets and resolutions are given in \autoref{tab:gen-times}. In some cases, generation times are lower for larger \numSupsegs, which is because \sensitivity~search converged faster.

While bisection search for \sensitivity~requires each ray to pass through the volume multiple times, it generates more accurate VDIs with each ray adaptively finding a near-optimal \sensitivity, instead of using one manually-tuned value across all rays.
Importantly, automatic \sensitivity~determination enables remote VDI generation without user intervention and streaming to a display client.

\begin{table}[]
\resizebox{\columnwidth}{!}{%
\begin{tabular}{l|l|l}
\hline
Dataset & \numLists=\shdres~& \numLists=\fhdres~\\ \hline \hline 
Kingsnake & 0.21 / 0.23 / 0.25 & 0.40 / 0.42 / 0.46  \\ 
Rayleigh-Taylor & 0.43 / 0.40 / 0.55 & 0.73 / 0.63 / 0.80 \\ 
Richtmyer-Meshkov & 0.75 / 0.82 / 0.82 & 1.55 / 1.36 / 1.36 \\ \hline 
\end{tabular}%
}
\caption{Wall-clock times in seconds to generate a single VDI with \numSupsegs = 15 / 20 / 30 for the datasets from \autoref{tab:datasets}. The camera is rotated about the data in 10\textdegree~increments, and means over 30 camera positions are reported. See \autoref{fig:image_comparisons} for transfer functions.}
\label{tab:gen-times}
\end{table}

\section{VDI Rendering by Raycasting}\label{sec:main-alg}

For each pixel in the viewport, we cast a ray into the VDI. The ray passes through supersegment lists \supseglist{}, searches for supersegments \superseg{}{} within them, and calculates the intersection length with each \superseg{}{}. It uses this to adjust the color contribution from each supersegment before accumulating them by alpha compositing. 

\subsection{Ray Traversal Through Lists}
The traversal of a ray is illustrated in \autoref{fig:raycasting}. The start and end points for the ray marching are determined by the intersection of ray with the viewport that was used to create the VDI and the bounding box of the volume in the scene. In \autoref{fig:raycasting}, for example, the ray marching begins at $t_\mathrm{near}$ and ends at $t_\mathrm{far}$.

\begin{figure}
 \centering
 \includegraphics[width=\columnwidth]{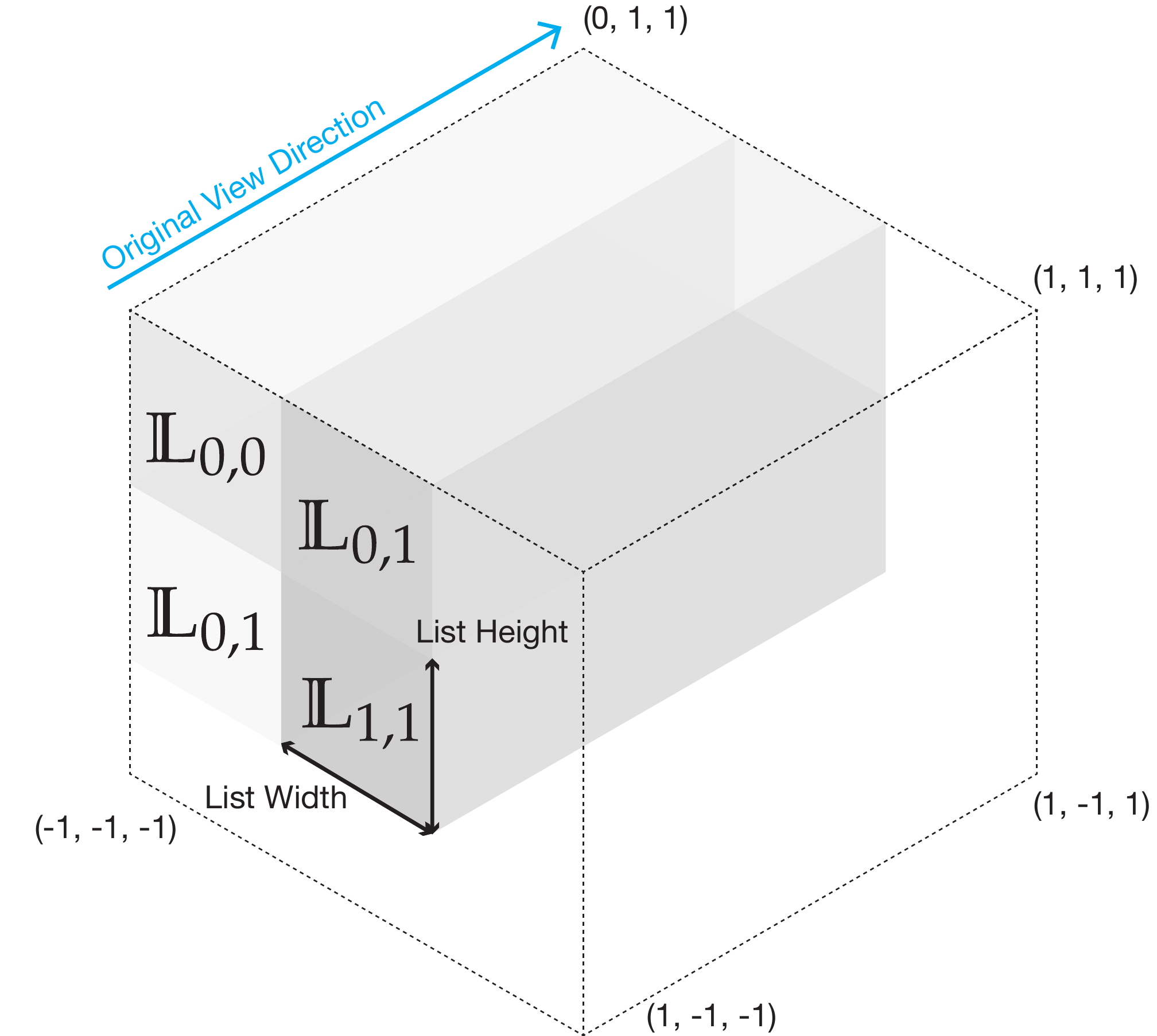}
 \caption{The supersegment lists \supseglist{} form a regular 2D grid of cuboids in the NDC space of \vieworig. In the \texttt{OpenGL} convention used in this figure, NDC range from -1 to 1 along all axes. The width and height of each cell (List Width and List Height) are therefore $2/w$ and $2/h$, respectively, where $w$ and $h$ are the $x$- and $y$-resolutions of the viewport used for generating the VDI.}
 \label{fig:ndc_grid}
\end{figure}

To simplify traversing through \supseglist{} and determining intersection points with \superseg{}{}, the calculations are done in the perspective-deformed Normalized Device Coordinate (NDC) space of \vieworig. In world space, \superseg{}{} and \supseglist{} form pyramidal frustums. In perspective NDC space, they are cuboids (\autoref{fig:clipspace}), with \supseglist{} forming a regular 2D grid as illustrated in \autoref{fig:ndc_grid}. A ray then traverses this grid using the fast voxel traversal algorithm by Amanatides and Woo \cite{amanatides}. With only two floating-point and two integer additions, and one floating-point and one integer comparison per iteration, the \supseglist{} intersected next is determined, along with the intersection points with a given \supseglist{i}, which are then used to search for \superseg{i}{j} within \supseglist{i}.

\subsection{Supersegment Search with Spatial Smoothness}\label{sec:search}

For each intersected \supseglist{}, we find the \superseg{}{} that cover the region between the entry and exit points of the ray. Once the first of these \superseg{}{} is determined (if any), the next is found by adjacency search, since the \superseg{}{} within \supseglist{} are sorted by their position. The algorithm only needs to check the next or preceding index, depending on the direction of the ray, which is defined by the sign of the scalar product between the ray and the original ray vector in world space.

\begin{algorithm}
  \caption{Find the first supersegment intersected in \supseglist{i}}
  \label{alg:first_supseg}
\begin{algorithmic}
\IF{\supseglist{i} is the first list intersected}
    \RETURN binary search from 0 to \numSupsegs
\ENDIF
\STATE $p \leftarrow$ index of last supersegment intersected in previous \textit{list}
\IF{\superseg{i}{p} is the first intersected supersegment in \supseglist{i}}
    \RETURN $p$
\ELSE
    \IF{ray position is behind \superseg{i}{p}}
    \RETURN binary search between $0$ and $p-1$
    \ELSE
    \RETURN binary search between $p+1$ and \numSupsegs
    \ENDIF
\ENDIF
    
\end{algorithmic}
\end{algorithm}

\begin{figure}
 \centering
 \includegraphics[width=\columnwidth]{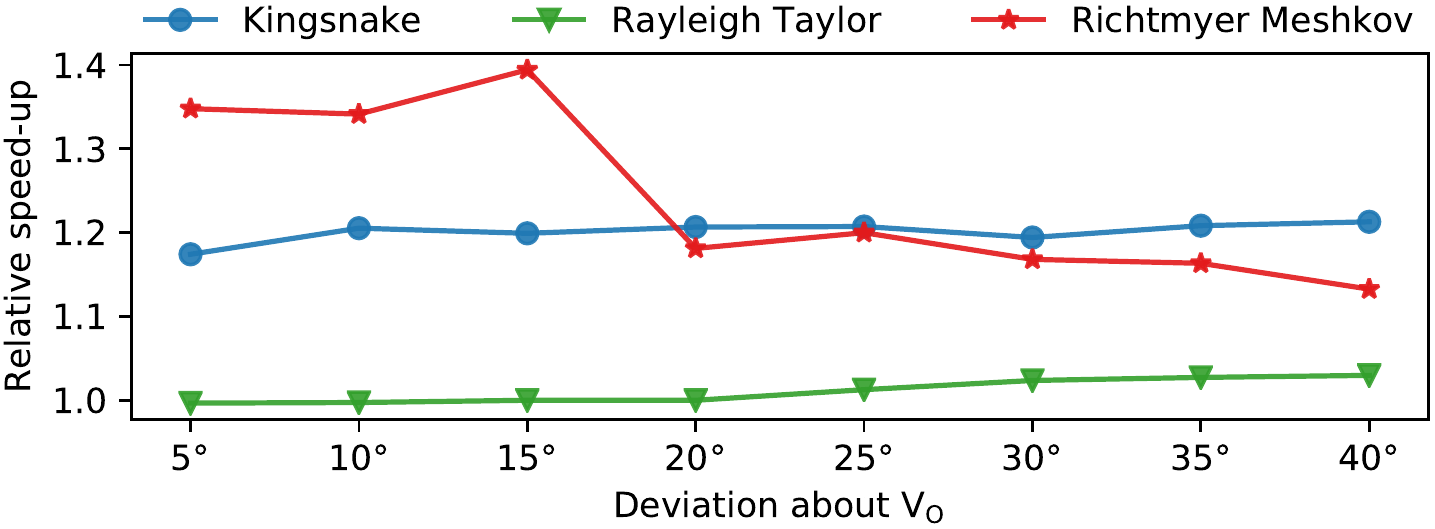}
 \caption{Speed-up in VDI rendering frame rates when using Alg.~\ref{alg:first_supseg} compared to, in each case, the best-performing among linear or binary search for the first \superseg{}{}. A VDI generated from \vieworig~is rendered at rotating viewpoints for three different datasets (symbols).}
 \label{fig:search_speedup}
\end{figure}

The procedure to determine the first intersected supersegment in a list is detailed in Alg.~\ref{alg:first_supseg}. The algorithm minimizes memory accesses by leveraging spatial smoothness in the volume data: neighboring \supseglist{} are created from rays that pass nearby in the data, and they are thus likely to have similar sized \superseg{}{}. The index $p$ of the last supersegment \superseg{h}{p} intersected in the previous list \supseglist{h} is therefore used as an initial guess for the first \superseg{}{} in the current list \supseglist{i}. If no \superseg{}{} was intersected in \supseglist{h}, $p$ is the index of the closest \superseg{}{} from the previous exit point. In \autoref{fig:clipspace}, for example, when the ray enters \supseglist{2}, it first tests for intersection with \superseg{2}{2}, since $2$ was the last index intersected in \supseglist{3}. The algorithm is further analyzed in the Supplement.

\autoref{fig:search_speedup}~reports the relative speed-up in VDI rendering frame rates when using Alg.~\ref{alg:first_supseg} compared to the respective best of binary or linear search for the first supersegment in a list. Baseline binary search is initialized with the middle \superseg{}{} in the \supseglist{}, and linear search with the first \superseg{}{}. 
In all cases, subsequent \superseg{}{} are found by adjacency search; the only difference is in how the first \superseg{}{} is found. Alg.~\ref{alg:first_supseg} yields up to 40\% better frame rates.

\begin{figure}
 \centering

 \includegraphics[width=\columnwidth]{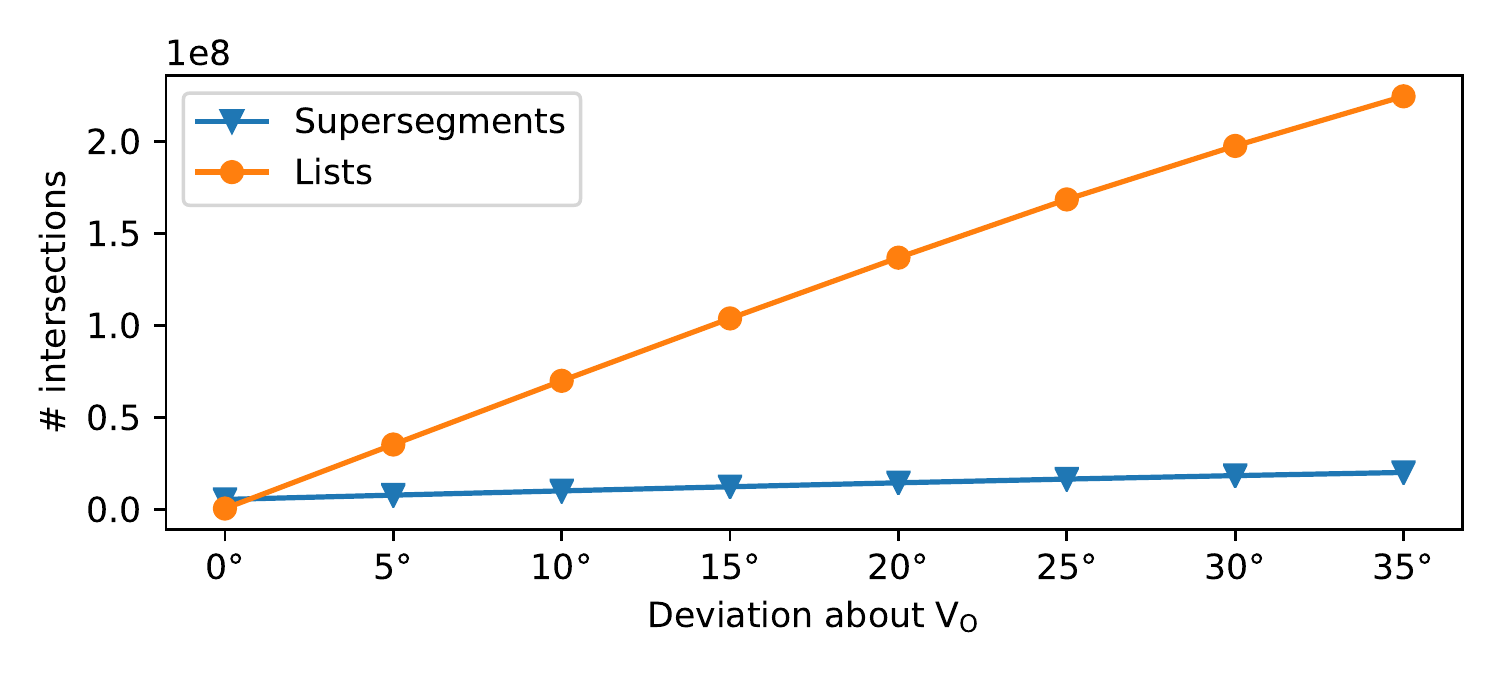}
 \caption{The total number of supersegment lists \supseglist{} and supersegments \superseg{}{} (symbols) intersected by all rays for different \viewnew~around \vieworig~for an \numLists=\fhdres, \numSupsegs=20 VDI of the Kingsnake dataset.}
 \label{fig:lists_vs_supsegs}
\end{figure}

For each intersected supersegment, the opacity accumulated by the ray needs to be adjusted by the intersection length of the ray with the supersegment, as \cite{correction}:
\begin{equation}\label{eq: correction}
\widetilde{\alpha} = 1 - (1-\alpha)^l
\end{equation}
where $\widetilde{\alpha}$ is the adjusted opacity, $\alpha$ is the opacity stored in the supersegment, and $l$ is the intersection length. Intersections are computed in NDC space, but adjusting opacity requires the intersection length $l$ in world space. This needs two additional matrix-vector multiplications to convert both intersection points to world space. Raycasting in the NDC space of \vieworig, in comparison to a world (or view) space technique, such as the one proposed by Lochmann et al.~\cite{novelview}, therefore optimizes intersections with supersegment lists, but has a higher cost for each supersegment intersected. \autoref{fig:lists_vs_supsegs} plots the number of supersegment lists \supseglist{}~and supersegments \superseg{}{}~intersected during VDI rendering at various viewpoint deviations around \vieworig. The number of \supseglist{}~traversed is larger than the number of \superseg{}{}~intersected, since for the majority of lists no supersegments are found. While the difference is particularly stark for the sparse Kingsnake dataset plotted here, this is true for more dense datasets, too. The difference between the number of \superseg{}{} and \supseglist{} intersected also grows with viewpoint deviation due to the anisotropic shape of the VDI. Early ray termination limits the number of \superseg{}{} intersected, while there is no limit on the number of \supseglist{}~intersected. The present strategy thus optimizes the costliest part of the rendering, which is the traversal of \supseglist{}.

\subsection{Empty-Space Skipping}\label{empty-space}
\begin{figure}
 \centering
 \includegraphics[width=\columnwidth]{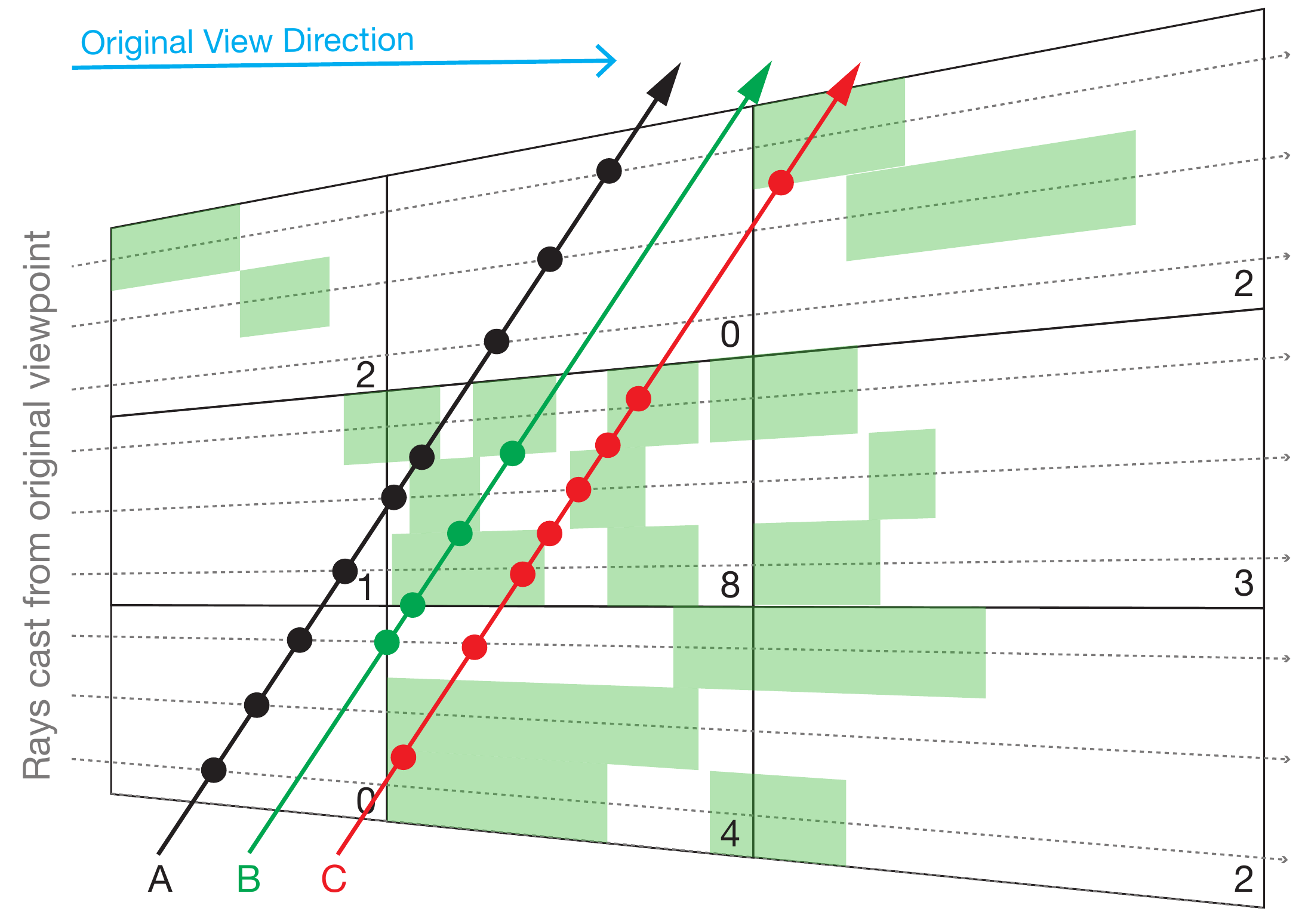}
 \caption{The grid data structure used for empty-space skipping and for preview rendering. Numbers in lower right corners are the values stored in each cell, which indicate the number of supersegments intersecting that cell. The rays of different color illustrate traversal strategies: Ray A is the base raycasting algorithm, ray B skips empty cells, and ray C subsamples the VDI for preview rendering. Dots indicate points at which the rays query the VDI to search for a supersegment in a list.}
 \label{fig:accel_grid}
\end{figure}

The performance of the present method depends on the number of memory accesses along a ray.
We optimize this by an acceleration data structure on top of the VDI to skip empty regions. The VDI implicitly contains the information required to skip empty space within a \supseglist{}, since \superseg{}{} are sorted by front and back depths defining their position in \supseglist{}. However, this cannot be queried based on the ray position when moving from one \supseglist{} to the next. Consider the example in Fig.~\ref{fig:accel_grid}, where Ray A must sample each \supseglist{i} at least once, even in empty areas, to determine that no \superseg{i}{j} in those \supseglist{i} is intersected. 

We therefore supplement the VDI with a grid acceleration data structure that stores the number of \superseg{}{} overlapping with each grid cell (black numbers in Fig.~\ref{fig:accel_grid}). This count, rather than just a binary indicator, is needed for preview rendering as explained in the next section.
When a ray hits a cell storing a 0, it jumps to the other end of that cell. This way, Ray B skips several \supseglist{} in the empty regions covered by the first and third cell it traverses. Querying the grid data structure requires one memory access per cell.

The grid cells have a constant depth extent in view space. In NDC space, this corresponds to cells that are larger toward the near plane, and smaller toward the far plane. Each cell spans an equal number of \supseglist{} along both dimensions of the viewing plane. The depth of a cell in view space is larger than its width or height, due to the anisotropic nature of a VDI (there are fewer \superseg{i}{j} in \supseglist{i} than there are \supseglist{i}).

The 3D regular grid could be replaced by hierarchical grids that provide better empty-space skipping performance, such as an octree \cite{octree} or SparseLeap \cite{sparse_leap}. However, we chose the current structure since it can be created during VDI generation with each \superseg{i}{j} generated triggering an atomic add on the appropriate grid cell.

\subsection{Dynamic Subsampling for Preview Rendering}\label{sec:downsampling}

The performance of the proposed raycasting algorithm depends on the number of memory accesses and calculations made by the rays as they traverse the VDI. Around the original view direction (\vieworig), high frame rates can be achieved as the VDI is compressed along \vieworig. For larger viewpoint deviations, rendering performance reduces, as larger portions of the rays go along one of the view plane dimensions. 
As illustrated in \autoref{fig:lists_vs_supsegs}, the number of intersected supersegment lists \supseglist{} increases with increasing viewpoint deviation. Each \supseglist{} intersected in a non-empty cell requires memory accesses to search for \superseg{}{}.

In the proposed remote visualization application, new VDIs are generated when the user's viewpoint changes. If large deviations from \vieworig~occur at the display client faster than a new VDI can be generated and streamed, however, the old VDI is used to bridge the time, trading off quality. In order to maintain frame rates, this uses dynamic sub-sampling for preview rendering.

\begin{figure}
 \centering
 \includegraphics[width=1.15\columnwidth]{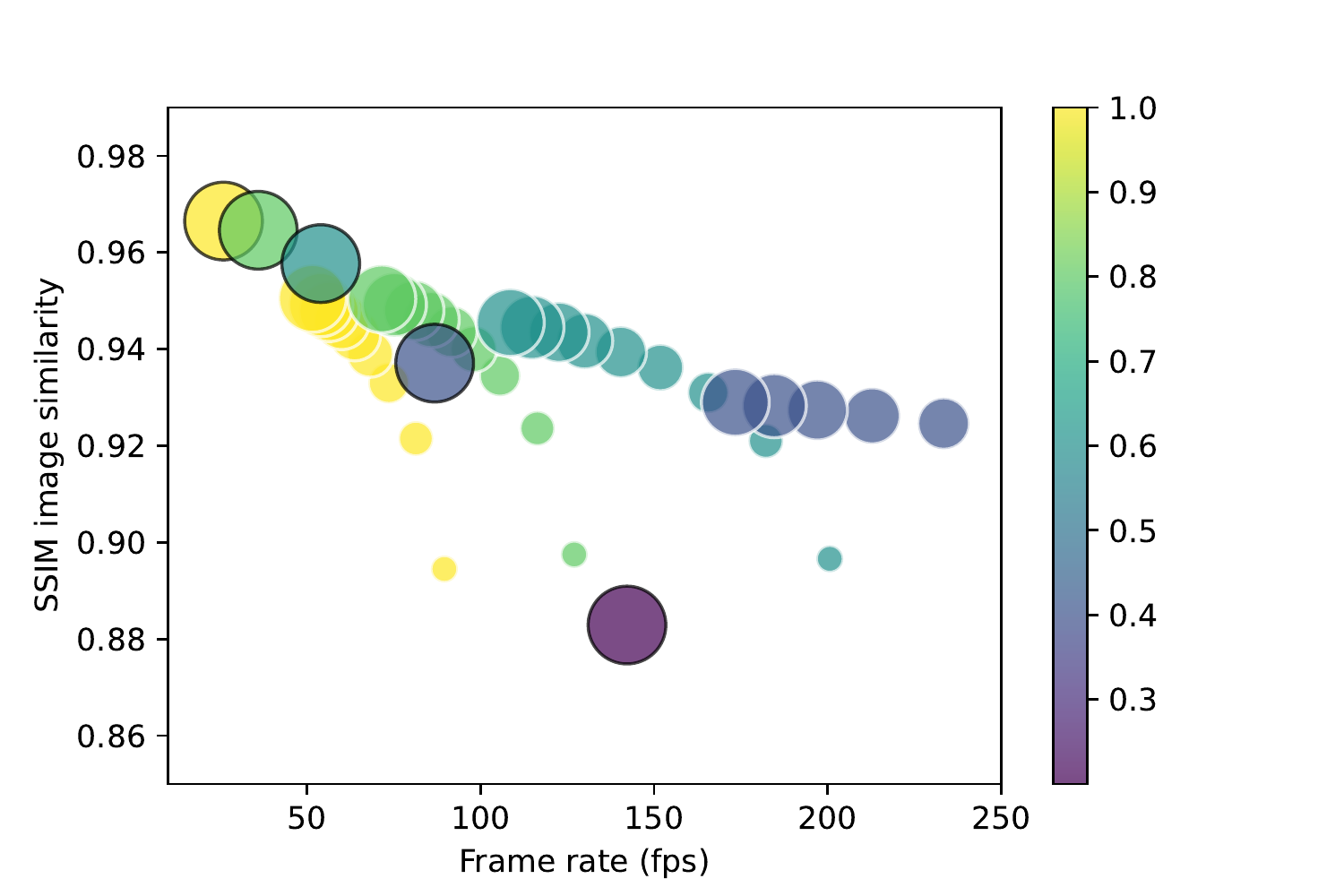}
 \caption{Performance (fps) and quality (SSIM w.r.t.~DVR) of adaptively sub-sampling VDI rendering along image dimensions (\downimage) and along the ray (\downray). Color is used to represent \downimage, smaller circle radii indicate smaller \downray. Circles with black outlines use full-resolution rendering along the ray. Display resolution is always 1920\by1080. Images rendered with \downimage \textless 1.0 are upsampled for display using bilinear interpolation.}
 \label{fig:downsampling}
\end{figure}

One way to achieve preview rendering would be to decrease the number of rays cast, followed by upsampling to the desired display resolution. Another way is to sub-sample along rays. We analyze the influence of both on VDI rendering performance and  quality. Quality is measured using the SSIM \cite{ssim} w.r.t.~fully resolved DVR. Higher SSIM indicate better rendering quality, with 1.0 indicating identical images. \autoref{fig:downsampling} shows the results for a \numLists=\fhdres, \numSupsegs=20 VDI generated on the Richtmyer-Meshkov dataset, rendered at 30\textdegree~from \vieworig. The VDI is the same in all cases.

The largest circles of each color, highlighted with a black outline, indicate the quality and performance obtained at various levels of downsampling (\downimage) in image space, i.e., by decreasing number of rays. \downimage~is indicated by color, with 1.0 indicating a full-resolution rendering. 
At \downimage=1.0 (yellow), the VDI renders at 45 fps. Frame rates increase when decreasing \downimage, but only slowly. To achieve a frame rate of $\approx$150 fps for example requires \downimage=0.2, which generates an image with SSIM=0.88. This is significantly worse than the SSIM of 0.97 for \downimage=1.0. 

We therefore propose to additionally sub-sample the rendering along the view ray, limiting the number of memory accesses and allowing higher values of \downimage~for the same set frame rate. In order to sample the VDI according to its information content, we use the acceleration data structure (\autoref{empty-space}) to distribute samples along the ray proportional to the number of \superseg{}{} in each grid cell. The samples are then uniformly thinned by a factor \downray. 
Rendering is further simplified by not calculating \superseg{}{} intersections. As the ray marches, it simply queries which \superseg{i}{j} a given sample point lies within, if any, and obtains the color from \superseg{i}{j}. Length-based opacity correction (Eq.~\ref{eq: correction}) is approximated using the distance from the previous sample. Since \superseg{i}{j} intersections are not computed, sampling is done in world space. The \supseglist{i} of a sample is found from its $x$- and $y$-coordinates in NDC space. The \superseg{i}{j} in \supseglist{i} is found using the algorithm from \autoref{sec:search}.

The number of samples within each cell of the acceleration grid is found by multiplying the sampling rate \downray~with the intersection length of the ray with the cell and with the number of \superseg{}{} in the cell, which is stored in the grid (\autoref{empty-space}). Samples are then placed at regular intervals along the ray within each cell. 
This amounts to adaptive sampling, as empty cells are not sampled and regions covered by more \superseg{}{}~are sampled more finely. Ray C in Fig.~\ref{fig:accel_grid} shows an example.

Adaptively sub-sampling the VDI enables rendering higher-quality images with larger  \downimage~for the same frame rate, as shown in \autoref{fig:downsampling}. Circles with white outline represent adaptive sub-sampling with smaller values of \downray~shown by smaller radii. The same $\approx$150 fps can now be achieved with significantly higher SSIM of 0.94 with \downimage=0.6 in image space and \downray=0.14 along the rays.

Sub-sampling the VDI for preview rendering requires choosing \downimage~and \downray~to obtain good image quality for a given target frame rate. We analyzed several datasets at multiple view configurations (see Supplement for details) and, while the results are qualitatively similar to \autoref{fig:downsampling}, the values at which a set of parameters becomes better than another one vary widely. The present remote visualization application (\autoref{sec:application}) therefore uses a dynamic PI (Proportional-Integral) controller to modulate \downimage, while allowing the user to choose the target frame rate and \downray.  

\section{Interactive Remote Visualization using VDIs}\label{sec:application}

Using the described methods to adaptively generate VDIs and to render them efficiently, we present the design of an application that enables remote visualization of large data volumes at high frame rates. VDIs are generated on a  server and streamed to a display client where they are rendered. Compared to remote DVR and streaming of rendered images, the proposed approach affords better interactive responsiveness, since the VDI is rendered locally on the client,  bypassing network latency for small viewpoint changes.

Each time the user changes the camera viewpoint on the display client, the camera pose is communicated to the server. As soon as the server finishes generating a VDI, it reads the most recent camera pose from the network. If the pose is different from the one of the previous VDI, or if the data themselves changed, a new VDI is generated. 
The data can change, e.g., if the server provides live {\insitu} visualization of a  simulation. The VDI can optionally be compressed before transmission to the client. Compression and streaming of the VDI are done asynchronously, interleaved with the generation of the next VDI, using separate CPU threads. Streaming of VDIs and communication of viewpoint changes is implemented using the networking library ZeroMQ (\href{https://zeromq.org/}{see zeromq.org}).

At the display client, receiving, decompressing, and uploading the VDI to the GPU are done asynchronously with VDI rendering on the GPU. Double buffering is used on the GPU to ensure that the previous VDI can be rendered while a new one is uploaded. The client always receives only the most recent VDI generated by the server. Near the viewpoint from which the VDI was generated (\vieworig), full-resolution VDI rendering is performed, maintaining high image quality at high frame rates. If frame rates drop below a user-set minimum at larger deviations from \vieworig, rendering transparently switches to adaptive sub-sampling for preview rendering until the new VDI is received from the server. At that moment, full-resolution rendering resumes. Note, however, that rendering of the new VDI need not begin from the viewpoint of generation, as the user may already have changed the camera position meanwhile. Future extensions could explore techniques that predict the user's camera movements to speculatively generate a VDI from the predicted viewpoint.

We implemented this remote visualization application in the open-source visualization library \scenery. The source code is available under the BSD license at \href{https://github.com/scenerygraphics/scenery}{github.com/scenerygraphics/scenery}. The implementation uses the high-performance Vulkan graphics API, enabling it to run across GPUs from different manufacturers.
Both rendering and generation of VDIs are implemented using compute shaders. For work distribution in the compute shaders, a local workgroup size of $16\times16$ is used, i.e., the screen space is divided into 2D blocks of that size. Each ray within a block corresponds to a thread on the GPU and a single pixel on screen. Each VDI consists of two floating-point textures: one for storing color and opacity of supersegments (type \texttt{RGBA32F}) and one for the depth of the supersegments (type \texttt{R32F}), with both front and back depth stored within.

\section{Evaluation}


\begin{table}[]
\centering
\begin{tabular}{|l|l|lll|}
\hline
\multirow{2}{*}{Dataset} &
  \multirow{2}{*}{\begin{tabular}[c]{@{}l@{}}VDI resolution\\ with $\numSupsegs=20$\end{tabular}} &
  \multicolumn{3}{l|}{\viewnew} \\ \cline{3-5} 
                           &                            & \multicolumn{1}{l|}{10º}            & \multicolumn{1}{l|}{20º}          & 40º          \\ \hline \hline
\multirow{4}{*}{Engine}    & \multirow{2}{*}{$512\times512$}   & \multicolumn{1}{l|}{28}            & \multicolumn{1}{l|}{28}           & 27           \\ \cline{3-5} 
                           &                            & \multicolumn{1}{l|}{\textbf{1174}} & \multicolumn{1}{l|}{\textbf{428}} & \textbf{291} \\ \cline{2-5} 
                           & \multirow{2}{*}{$1024\times1024$} & \multicolumn{1}{l|}{9}             & \multicolumn{1}{l|}{9}            & 9            \\ \cline{3-5} 
                           &                            & \multicolumn{1}{l|}{\textbf{454}}  & \multicolumn{1}{l|}{\textbf{80}}  & \textbf{60}  \\ \hline
\multirow{4}{*}{Kingsnake} & \multirow{2}{*}{$512\times512$}   & \multicolumn{1}{l|}{31}            & \multicolumn{1}{l|}{31}           & 30           \\ \cline{3-5} 
                           &                            & \multicolumn{1}{l|}{\textbf{359}}  & \multicolumn{1}{l|}{\textbf{298}} & \textbf{234} \\ \cline{2-5} 
                           & \multirow{2}{*}{$1024\times1024$} & \multicolumn{1}{l|}{11}            & \multicolumn{1}{l|}{10}           & 10           \\ \cline{3-5} 
                           &                            & \multicolumn{1}{l|}{\textbf{441}}  & \multicolumn{1}{l|}{\textbf{50}}  & \textbf{44}  \\ \hline
\multirow{4}{*}{\begin{tabular}[c]{@{}l@{}}Rayleigh-\\ Taylor\end{tabular}} &
  \multirow{2}{*}{$512\times512$} &
  \multicolumn{1}{l|}{20} &
  \multicolumn{1}{l|}{18} &
  18 \\ \cline{3-5} 
                           &                            & \multicolumn{1}{l|}{\textbf{710}}  & \multicolumn{1}{l|}{\textbf{589}} & \textbf{401} \\ \cline{2-5} 
                           & \multirow{2}{*}{$1024\times1024$} & \multicolumn{1}{l|}{5}             & \multicolumn{1}{l|}{5}            & 5            \\ \cline{3-5} 
                           &                            & \multicolumn{1}{l|}{\textbf{226}}  & \multicolumn{1}{l|}{\textbf{167}} & \textbf{124} \\ \hline
\end{tabular}
\caption{Comparing the performance (in fps) of our approach (bold) for rendering VDIs with the rasterization-based technique from \cite{Frey}.}
\label{table:comparison_rast}
\end{table}

We evaluate our implementation of the present algorithms on the real-world datasets from \autoref{tab:datasets}. 
For evaluation, VDIs were generated from one or more viewpoints (\vieworig) and rendered at one or more new viewpoints \viewnew. The \viewnew~were rotations of the camera about the dataset center with the camera always facing the dataset center.
Unless otherwise stated, results are reported as frame rates, i.e., the inverse of the frame time at a given camera pose.

\subsection{Comparison with other VDI Rendering Techniques}

\begin{figure*}[]
    \centering
    \begin{subfigure}[t]{0.48\textwidth}
        \centering
        \includegraphics[height=1.9in]{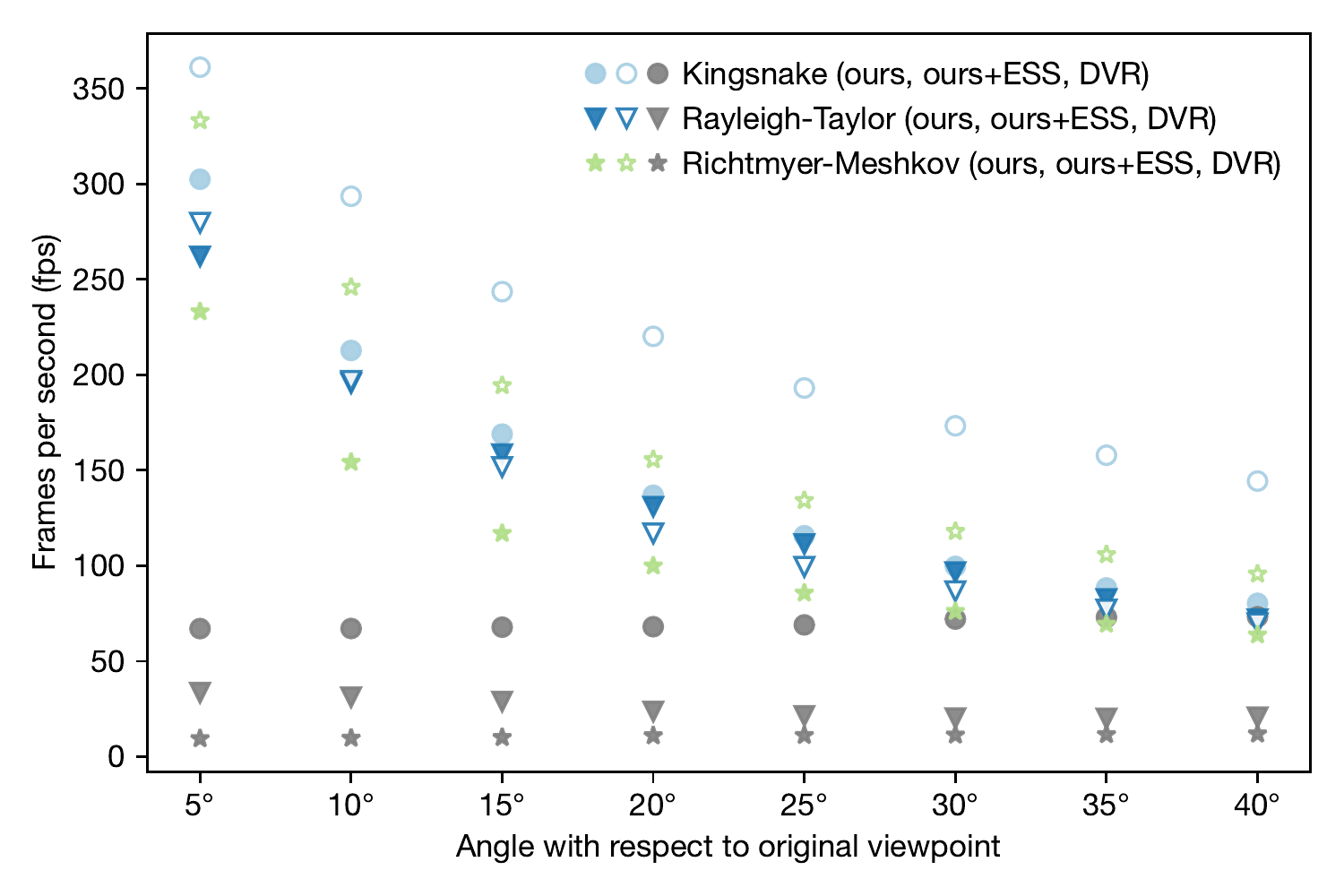}
        \caption{Rendering frame rates for \shdres, \numSupsegs=20, VDIs.}
        \label{fig:benchmark_720}
    \end{subfigure}%
    ~ 
    \begin{subfigure}[t]{0.48\textwidth}
        \centering
        \includegraphics[height=1.9in]{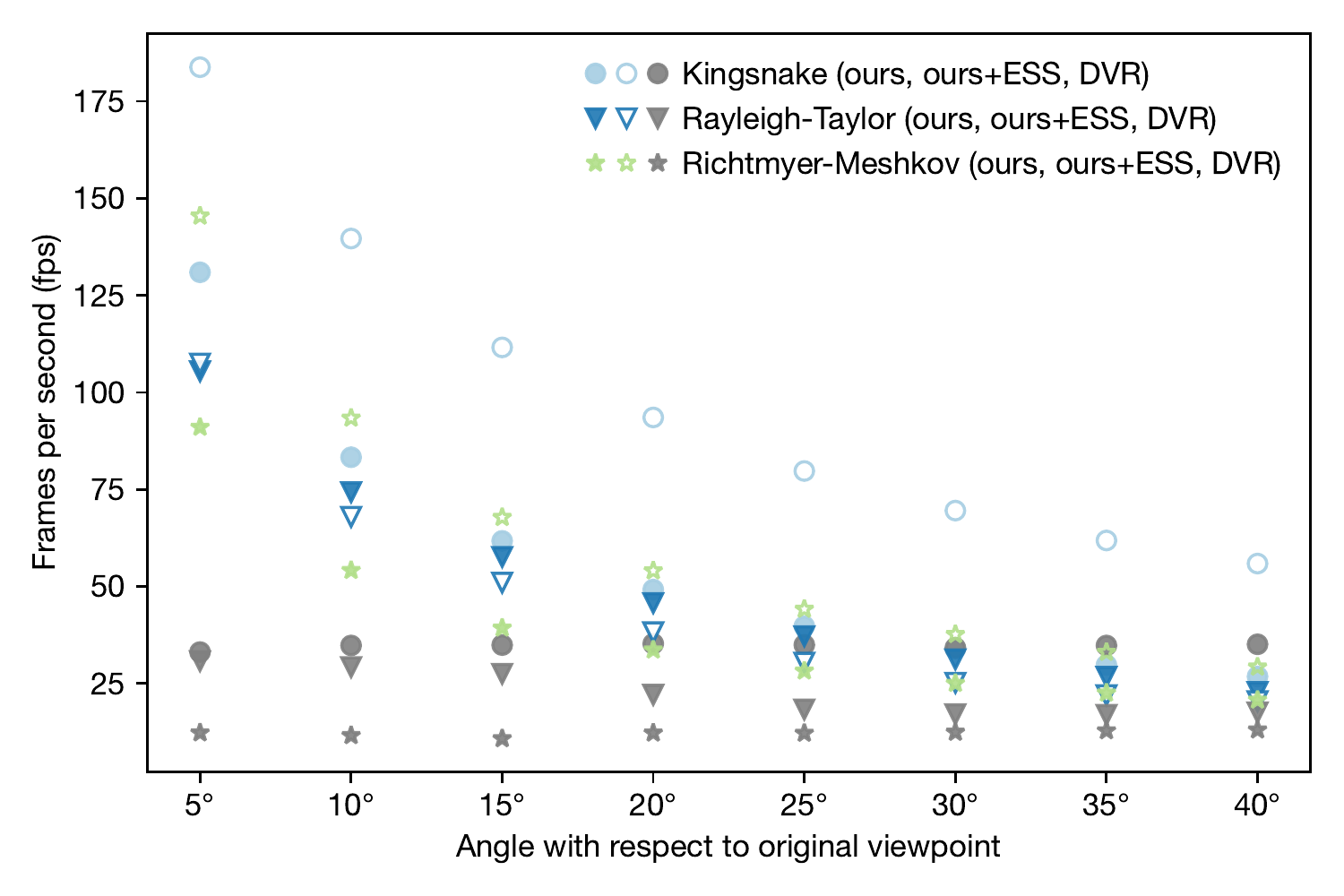}
        \caption{Rendering frame rates for \fhdres, \numSupsegs=20, VDIs.}
        \label{fig:benchmark_1080}
    \end{subfigure}
    \caption{VDI rendering frame rates without (``ours'', blue/green filled symbols) and with empty-space skipping (``ours+ESS'', blue/green open symbols) for different datasets, compared with direct volume rendering (``DVR'', gray symbols) from the same viewpoint.}
    \label{fig:vdi_vs_volume}
\end{figure*}

\begin{figure} 
        \centering
        \includegraphics[height=2in]{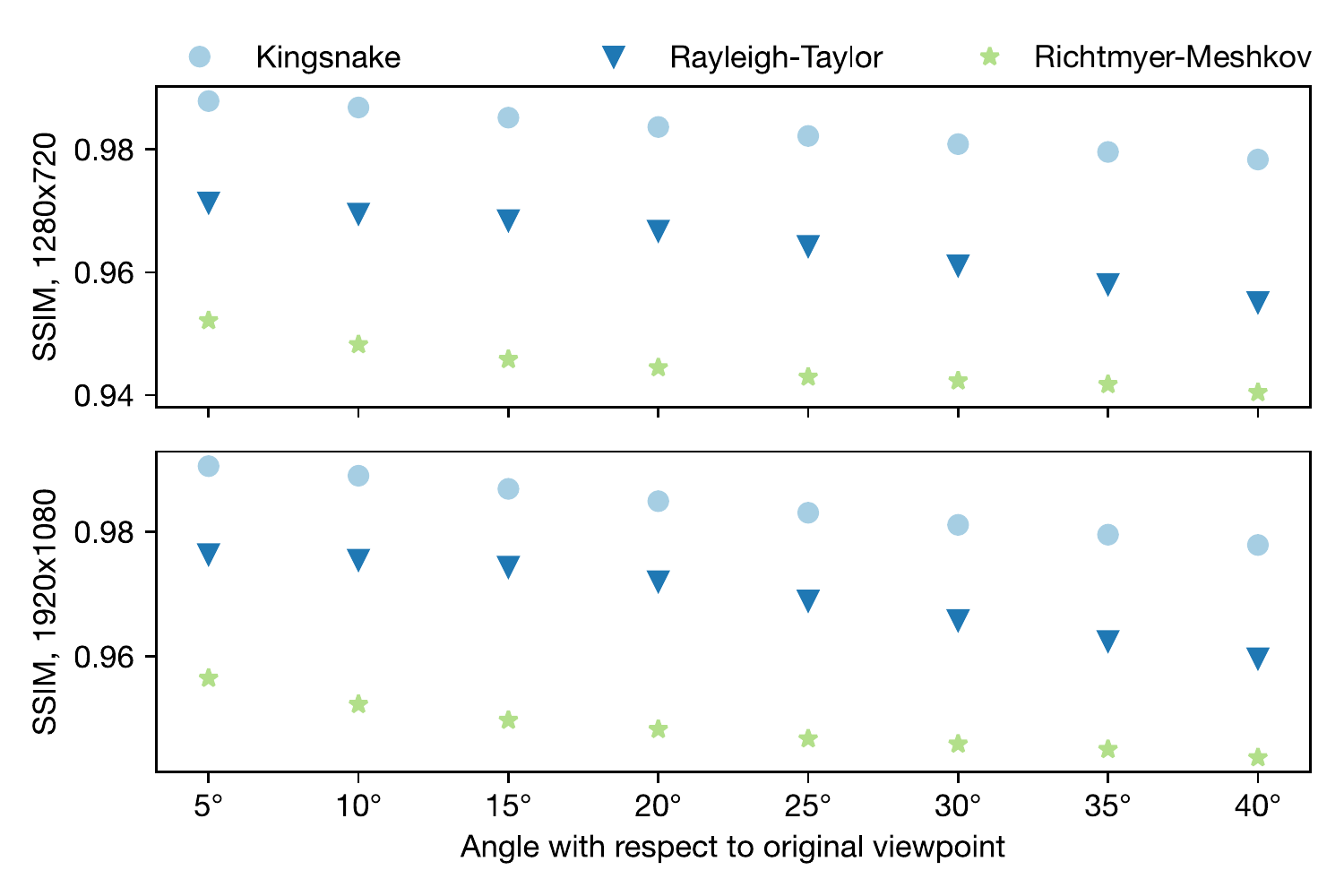}
        \caption{SSIM image similarity between VDI rendering and DVR from the same viewpoint for \shdres~(top) and \fhdres~(bottom) with \numSupsegs=20 for different datasets (symbols).}
    \label{fig:ssim_vdi}
\end{figure}

We first compare the present raycasting-based VDI rendering with the rasterization-based method of Frey et al.~\cite{Frey}. To ensure that both methods render identical VDIs, the VDIs were generated using the implementation of Frey et al.~and imported into our software for rendering. We also disabled empty-space skipping (\autoref{empty-space}) in our implementation, since the code by Frey et al. does not have it. \autoref{table:comparison_rast} reports the results. We could only use the three smaller datasets, and smaller viewport resolutions, for this comparison, due to technical limitations in the implementation by Frey et al.

In all cases, the present method outperforms the rasterization-based approach, often by more than an order of magnitude. For the implementation by Frey at al., performance is similar to that reported in the original 2013 paper, indicating that their technique of sorting supersegment lists \supseglist{} in front-to-back order for a given rendering viewpoint \viewnew, followed by $\alpha$-blending, does not substantially benefit from recent advances in GPUs.

The performance of Frey et al.~\cite{Frey} remains similar at all tested \viewnew. The present raycasting method, however, is faster at smaller viewpoint deviations. This is because raycasting can take advantage of the anisotropic view-dependent shape of the VDI, where rays can march quickly along supersegment lists, leading to higher frame rates near \vieworig. Another advantage of the raycasting approach can be observed for the dense Rayleigh-Taylor dataset, where early ray termination provides significant speed-ups not possible with object-space approaches like that of Frey et al. 
The Kingsnake dataset is challenging when disabling empty-space skipping, because it contains large empty regions. Still, the present method outperforms Frey et al., even without empty-space skipping.

The present raycasting approach is also more consistent in its memory requirement, since no geometry needs to be generated. The memory required by the code of Frey et al.~\cite{Frey} depends heavily on the dataset. For the dense Rayleigh-Taylor dataset at 1024\by1024 viewport resolution, Frey et al. generate 3473\,MB of geometry and for the sparse Kingsnake dataset only 349\,MB. Our code has a constant memory requirement of 480\,MB across all datasets for 1024\by1024 viewport resolution. In addition to providing faster rendering performance, our raycasting approach does not require creation of geometry which is a one-time per-VDI cost for Frey et al. The rendering quality produced by the two methods is the same, and therefore not compared.

We were unable to perform empirical performance comparisons with the raycasting technique proposed by Lochmann et al.~\cite{novelview} for a similar data structure, since their code is not publicly available. As explained in \autoref{sec:main-alg}, however, our technique of traversing the VDI in the NDC space of \vieworig~reduces the number of calculations compared to the strategy described by Lochmann et al., where traversal is performed in view space requiring intersections with pyramidal frustums (confirmed by original authors in personal communication). While Lochmann et al.~do not provide details on how supersegments are searched for within supersegment lists, our optimized search strategy  leveraging spatial homogeneity provides speed-ups of up to 40\% over binary search (\autoref{fig:search_speedup}). Lochmann et al.~also did not use empty-space skipping, likely losing further performance (\autoref{fig:vdi_vs_volume}).

\subsection{Comparison with Direct Volume Rendering}

Next, we compare the present approach with DVR. VDIs are generated from four different \vieworig, placed around the dataset at 90\textdegree~rotations, rendered at different deviations \viewnew~about each \vieworig, and compared with DVR from the same viewpoint. \autoref{fig:vdi_vs_volume} plots the mean frame rates over the four \vieworig~for two different viewport resolutions, \numLists=\shdres~(standard HD) and \numLists=\fhdres~(full HD). VDI frame rates are reported with and without empty-space skipping (ESS) for comparability with DVR, which does not use empty-space skipping.

At small \viewnew=5\textdegree, VDI rendering achieves significant speed-ups over DVR in the range of 4.5$\ldots$24.5$\times$ for standard HD and 3.5$\ldots$7.5$\times$ for full HD across datasets. VDI raycasting frame rates decrease for increasing \viewnew, as rays have to do more work due to the anisotropic view-dependent shape of the VDI. However, they remain higher than DVR at all \viewnew~except for the full HD rendering of the Kingsnake beyond 30\textdegree. The lower speed-ups for full HD resolution compared to standard HD are expected, since the sizes of the VDI and the rendering viewport both increase. Empty-space skipping mostly increased VDI rendering frame rates, particularly for the sparse Kingsnake and Richtmyer-Meshkov datasets, but occasionally reduced them slightly for the dense Rayleigh-Taylor dataset.

We also compared the quality of the images generated by VDI rendering with those from DVR. \autoref{fig:ssim_vdi} provides the results in terms of the SSIM (Structural Similarity Index Measure) \cite{ssim}, where higher values are better and 1.0 corresponds to identical images. \autoref{fig:image_comparisons} provides visual comparisons (see Supplement for full resolution images). Like frame rates, SSIM values are also higher for smaller \viewnew, as view rays are better aligned with the rays that generated the VDI. The reduction in rendering quality, however, is minor even at high \viewnew~of up to 40\textdegree. Similar results are observed when using the PSNR quality metric (see Supplement).

VDI rendering quality can further be increased by increasing \numSupsegs~at the cost of larger VDI size and reduced frame rates. Rendering frame rates, however, were found to reduce sub-linearly with increasing \numSupsegs, falling to between 0.61$\times$ and 0.89$\times$ when doubling \numSupsegs.

Videos can be found in the Supplementary material showing interactive visualization sessions using both DVR and VDI rendering on all datasets.

\subsection{Comparison with Remote Volume Visualization}

Finally, we compare the present VDI-based remote visualization system with remote DVR using NVENC video encoding for image streaming. For network streaming, our implementation compresses VDIs using the lossless LZ4 algorithm, which we found to provide the best trade-off between speed and compression in our benchmarks, in comparison with zstd and Snappy. LZ4 yields compressed VDIs of $\approx$100\,MiB for standard HD (\shdres) resolution and $\approx$225\,MiB for full HD (\fhdres). The corresponding uncompressed VDI sizes are 422\,MiB and 950\,MiB, respectively. In addition, the empty-space skipping data structure (2.5\,MiB for full HD) and the metadata about \vieworig~($\approx$200 Bytes) are also transmitted.

To evaluate the performance of the present system in its entirety, we compare with existing remote visualization functionality in \scenery, which offers video streaming with hardware-accelerated encoding and decoding using NVENC and CUVID, respectively. The volume data reside on a server with Nvidia GeForce RTX 3090 GPU, where DVR and VDI generation take place. The display client is a standard office workstation with an AMD Radeon RX 5700XT GPU, connected via 1\,Gbit/s Ethernet across rooms.

All camera viewpoint changes are applied synchronously at the client in order to ensure frame-to-frame comparability. Overall frame times are measured to estimate the ``motion-to-photon'' latency, i.e., the time between the user making a movement and the movement being fully reflected on the display. For the VDI, this is the rendering frame time of the VDI at the client. For remote DVR, it is the rendering frame time plus the streaming time, which includes the time to send the new camera pose to the server, encode the rendered frame, stream, and decode the frame at the client.

A camera path is pre-recorded, consisting of four phases that evaluate different modes of interactive visualization: Phases 1 (frames 0-500) and 4 (frames 1500-2000) show steady camera movements for data exploration; Phase 2 (frames 500-1000) consists of fast movements for navigation, and in Phase 3 (frames 100-1500) the camera zooms in on a point of interest and then out again. Screencasts of the  interactive sessions with DVR and VDI rendering are provided in the Supplement. \autoref{fig:remote}~reports the performance measurements for full HD resolution of the Richtmyer-Meshkov dataset.

Streaming time was found to add only a marginal overhead on DVR timings in our setup. While we were unable to measure decoding timings on the AMD Radeon RX 5700XT due to incompatibility with \textit{scenery}'s CUVID decoding, on the RTX 3090 we found decoding times $< 0.5$ ms per frame. NVENC encoding added a consistent overhead of approximately 5 ms. \autoref{fig:remote} also reports the frame numbers at which new VDIs become available for rendering, providing an estimate of the ``VDI latency'', which includes VDI generation, transmission, compression and decompression, as well as GPU upload. Note that this latency is hidden from the user through double buffering (\autoref{sec:application}).

\begin{figure} 
        \centering
        \includegraphics[width=0.97\columnwidth]{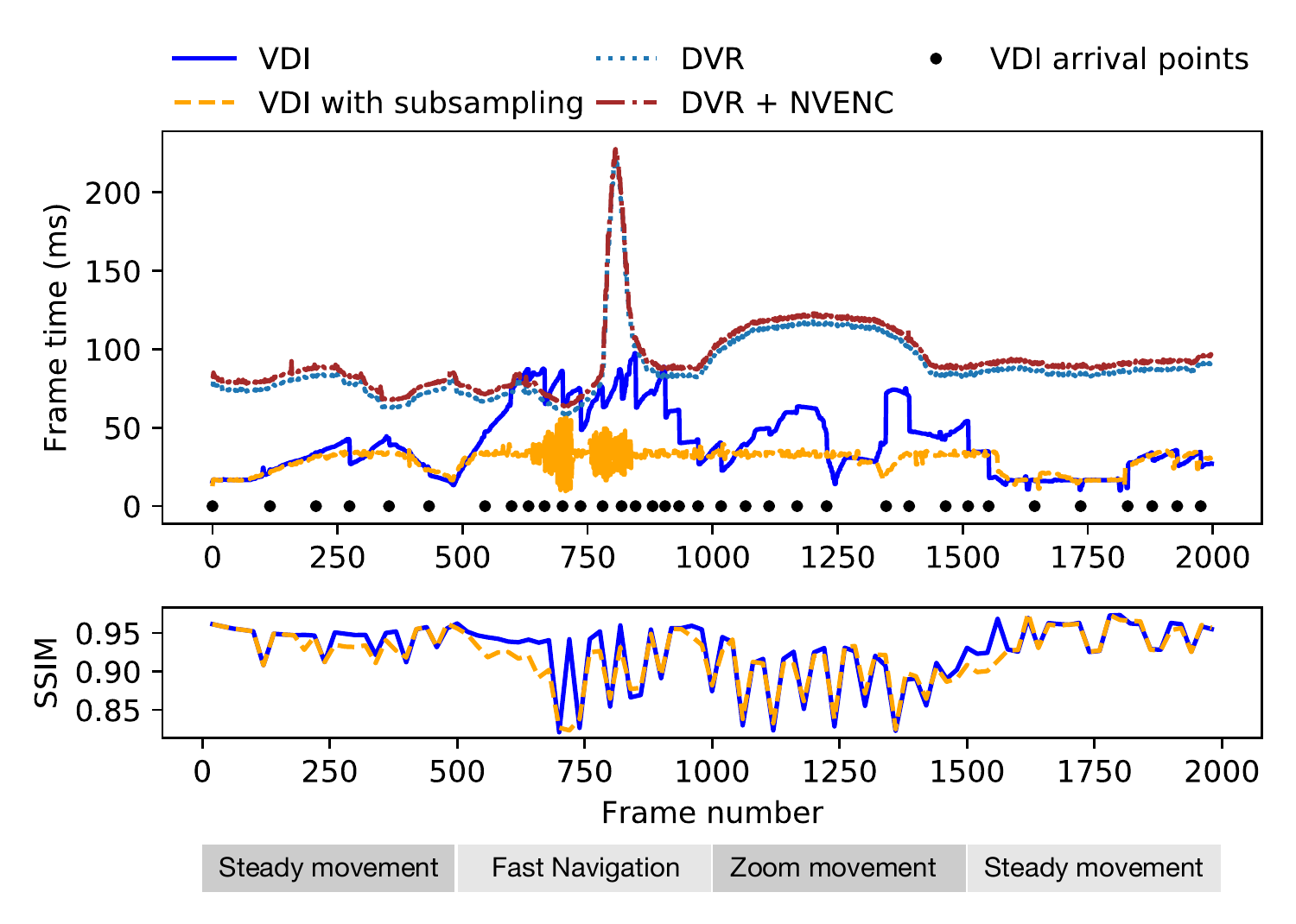}
        \caption{Top panel: end-to-end frame times for remote visualization of the Richtmyer-Meshkov dataset in full HD resolution using remote DVR (on an RTX 3090)
        and VDI rendering (on a Radeon RX 5700XT) over an interactive session of 2000 frames (see video in Supplement). Bottom panel: VDI rendering and VDI subsampling quality at every 20$^{\mathrm{th}}$ frame as SSIM w.r.t.~DVR images.}
        \label{fig:remote}
\end{figure}

During the steady camera movements of Phases 1 and 4, new VDIs that arrive from the server are rendered at not-too-distant viewpoints from \vieworig. This results in smooth interactive frame rates with frame-times significantly shorter than those achieved by remote DVR, while maintaining high rendering quality. VDI frame times are also significantly shorter than those of remote DVR during the zooming Phase 3. As the camera zooms back out towards the end of Phase 3, 
missing regions are visible in the VDI rendering. This is because the VDI generated from the zoomed-in viewpoint is still rendered, representing only data within its viewport. This leads to the drop in SSIM towards the end of Phase 3, until the new VDI for the zoomed-out viewpoint arrives.

The fast navigation in Phase 2 is challenging for VDI rendering, as large deviations from \vieworig~occur. This leads to increasing VDI frame times, eventually becoming comparable to those of DVR in this phase. 
Note, however, that VDI rendering runs on an AMD Radeon RX 5700XT, which is significantly less powerful ($0.24\times$ computational throughput, $0.44\times$ peak memory bandwidth) 
than the RTX 3090 used for DVR. 

We also evaluate the adaptive preview rendering strategy of \autoref{sec:downsampling} and demonstrate how the PI controller dynamically adjusts \downimage~to maintain a target frame rate. Adaptive sampling along the ray is manually activated during Phase 2 with \downray=0.3.
The PI controller is able to maintain a steady frame rate of 25 fps through most of the session, though jerkiness was observed during Phase 2. 

\section{Conclusions}
We have presented an efficient raycasting-based rendering method for VDIs and its use in remote visualization of large volume data. At its core is an efficient way of intersecting supersegment lists \supseglist{}~and supersegments \superseg{}{} by ray marching in the NDC space of \vieworig, where all \supseglist{}~are transformed from irregular pyramidal frustums to cuboids. These can then be traversed by voxel stepping \cite{amanatides}. We presented an efficient method for finding \superseg{}{}~within \supseglist{}~by leveraging spatial smoothness in the data. This increases frame rates by up to 40\% over the binary search baseline. We also presented a method for skipping empty space during raycasting.

The presented method significantly outperforms the rasterization-based VDI rendering by Frey et al.~\cite{Frey} (\autoref{table:comparison_rast}), which we found to not scale well to modern GPUs. Our raycasting approach also has a few inherent advantages: it does not require the creation of geometry, benefits from early ray termination, and better leverages the anisotropy of the VDI near \vieworig. While we were unable to present a direct comparison with the raycasting approach by Lochmann et al.~\cite{novelview} due to unavailability of their code, we argue that our \supseglist{}~traversal, \superseg{}{}~search, and empty-space skipping are likely to result in better performance.
We also found that our VDI rendering frame rates are significantly higher than DVR (\autoref{fig:vdi_vs_volume}) close to \vieworig, while providing high-quality approximations (\autoref{fig:ssim_vdi}).

Finally, we have shown how the present method can be used in a remote visualization application. In this context, we have proposed an extension to the VDI generation technique of Frey et al.~\cite{Frey}, where \sensitivity~values are adaptively determined for each ray, generating more accurate VDIs without manual parameter tuning. To perform preview rendering at large viewpoint deviations before the next VDI arrives, we proposed an adaptive sub-sampling along the ray, which we have shown to combine well with image-space sub-sampling to yield consistently high frame rates. Overall, the entire remote visualization application sustained higher frame rates than remote DVR (\autoref{fig:remote}).

A current limitation of our implementation is that the adaptive subsampling along the rays needs to be activated and tuned manually,
while subsampling in image space is dynamically controlled by a PI controller.
Future extensions could relax this limitation by exploring Multi Input Multi Output (MIMO) controllers. A limitation of the VDI data structure itself is that while it can model non-directional lighting, such as local ambient occlusion \cite{localAO}, directional effects, such as specular lighting, cannot be realized because the VDI stores classified color and opacity.

We see the presented rendering approach to VDIs as a step toward remote visualization of large volumes. The frame rates achieved by the present implementation are consistently higher than those of DVR and of other VDI rendering approaches. Further improvements, such as foveated rendering \cite{fovolnet,FAVR}, can potentially increase them even further in the future. 

\acknowledgments{
This work was supported by the Center for Scalable Data Analytics and Artificial Intelligence (ScaDS.AI) Dresden/Leipzig. This work was partially funded by the Center for Advanced Systems Understanding (CASUS), financed by Germany’s Federal Ministry of Education and Research (BMBF) and by the Saxon Ministry for Science, Culture and Tourism (SMWK) with tax funds on the basis of the budget approved by the Saxon State Parliament. We thank the 
University of Texas High-Resolution X-ray CT Facility (UTCT) for the Kingsnake dataset.
}

\bibliographystyle{abbrv-doi-hyperref}

\bibliography{references}
\end{document}




\maketitle
\section{Ray-Adaptive Generation of Supersegments}
Algorithm~\ref{alg:iterative} provides the pseudo-code for the proposed ray-adaptive generation of supersegments (Section 3 in the main text). Most of the parameters referenced in Alg.~\ref{alg:iterative} were introduced in the main text. The search terminates when the search space reduces below a small $\epsilon$, selecting the high end of the range as \sensitivity, which is guaranteed to produce less supersegments than \numSupsegs~and therefore prevent smearing, unless it would produce 0 supersegments (line 10). In our experiments, we set $\epsilon$ to $10^{-6}$.

\begin{figure*}
\centering
  \includegraphics[width=0.87\textwidth]{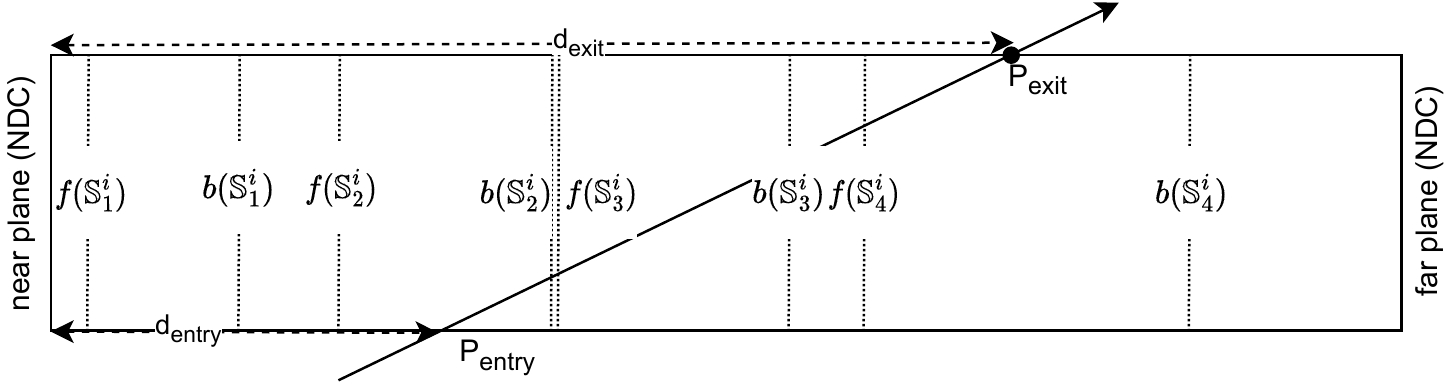}
 
  \caption{Illustrating the parameters involved in the search for the first intersected supersegment in \supseglist{i} (Alg.~\ref{alg:first_supseg}).}
  \label{fig:intersection}
\end{figure*}

\begin{algorithm}[]
\caption{Determining the supersegment termination parameter \sensitivity~using an iterative bisection search}
\begin{algorithmic}[1]
\STATE low $\gets$ 0
\STATE high $\gets$ $\sqrt{3}$
\STATE \sensitivity~$\gets$ 0.00001
\STATE firstIteration $\gets$ TRUE
\STATE found $\gets$ FALSE
\WHILE {!found} 
  \STATE n $\gets$ numSupersegmentsGenerated(\sensitivity)
  \IF {abs(high-low) $<$ $\epsilon$} 
    \STATE found $\gets$ TRUE
    \STATE \sensitivity~$\gets$ ((n$==0$) ? low : high)
  
  \ELSIF {n $>$ \numSupsegs} 
    \STATE low = high
  \ELSIF {n $<$ \numSupsegs$-\delta$}
    \STATE high = low
  \ELSE 
    \STATE found $\gets$ TRUE
  \ENDIF
  \IF {firstIteration} 
    \STATE firstIteration $\gets$ FALSE
    \IF {n $<$ \numSupsegs} 
      \STATE found $\gets$ TRUE
    \ENDIF
  \ENDIF
  \IF{!found} 
    \STATE \sensitivity~$\gets$ (low + high) / 2\;
  \ENDIF
\ENDWHILE
\end{algorithmic}
\label{alg:iterative}
\end{algorithm}

\section{Supersegment Search with Spatial Smoothness}

Algorithm \ref{alg:first_supseg} presents the pseudo-code for determining the first intersected supersegment in a list,  including optimizations to reduce memory latency on the GPU. Figure \ref{fig:intersection} illustrates some of the parameters used in the algorithm. 

As described in Section 4.2 of the main text, the algorithm minimizes memory accesses by leveraging spatial smoothness in the volume data. The index $p$ of the last supersegment \superseg{h}{p} intersected in the previous list \supseglist{h} is used as an initial guess for the first \superseg{}{} in the current list \supseglist{i}.

To find the first \superseg{}{}~in \supseglist{}, the distance of the entry point from the camera near-plane in NDC ($d_{entry}$) is compared to the \superseg{}{}~depths. To minimize the impact of memory latency, \back{i}{p}~and \back{i}{p-1}~are pre-fetched before comparisons begin (lines 8 -- 9). To minimize thread divergence within warps, a single line of code (line 29) calls the binary search function ``BinS" if \superseg{i}{p}~is not the first \superseg{}{} intersected, with appropriate function parameters pruning the search space. The number of comparisons are minimized by precomputing the interval, with respect to \superseg{i}{p}, in which $d_{entry}$~lies (line 10).

If \textit{p} is invalid (negative), the binary search range is set to search in the full list (lines 3 -- 4). If \superseg{i}{p} does not exist, $b_1$ is set to very far back (line 8), if \superseg{i}{p-1} does not exist, $b_0$ is set to very far in front (line 9). $d_{entry}$ is then compared to $b_0$ and $b_1$ to check is it falls between them (line 10). A value of 0 for \textit{interval} means the intersecting supersegment is behind \superseg{i}{p}, a value of 2 means the intersecting supersegment is in front of \superseg{i}{p}. In such cases the bin\_search\_start and bin\_search\_end are adjusted accordingly to reduce the search space for the binary search (lines 11 -- 16). A value of 1 means that the \superseg{i}{p} is potentially the first \superseg{}{}~in \supseglist{i}~intersected by the ray. Line 10 then checks whether \superseg{i}{p} actually exists. If so, superseg\_found is set to true, and the index $p$ is stored (lines {18 -- 22}). If \superseg{i}{p}~does not exist, a full binary search in \supseglist{i}~is used to find the intersecting supersegment (lines {22 -- 25}). A value of bin\_search\_end different from -1 indicates that a binary search must be performed to find the supersegment (lines {28 -- 29}). The last condition (lines {32 -- 34}) ensures that the determined supersegment really intersects the ray, If the front-depth of the determined supersegment is bigger than the exit distance of the ray ($d_{exit}$), this means the ray passes between supersegments without intersecting any, in which case $index$ is set to $-1$, and superseg\_found is set to false. Note that the conventions used here to explain this make the assumption that the angle between the intersecting ray and original ray is less than 90\textdegree, but the algorithm is equally applicable in both cases.

\begin{figure}
 \centering
 \includegraphics[width=\columnwidth]{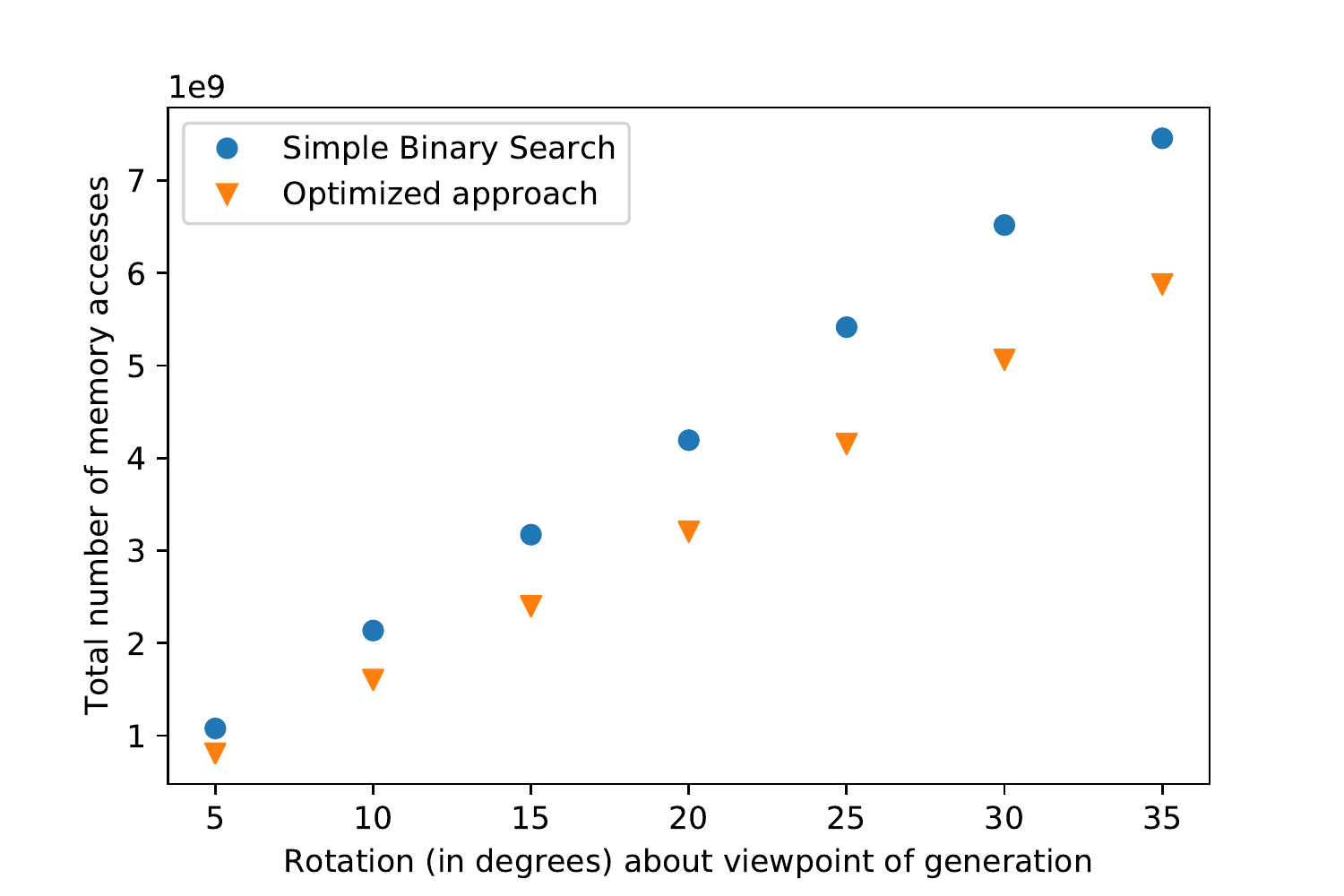}
 \caption{Comparing the the total number of memory accesses performed during VDI rendering with the first \superseg{}{} in \supseglist{} determined either using our optimized search, leveraging spatial smoothness, or using a simple binary search (symbols). Analysis was performed on a \fhdres~VDI with \numSupsegs=20, generated from the Kingsnake dataset.}
 \label{fig:opt_vs_binary}
\end{figure}

Figure~\ref{fig:opt_vs_binary} provides a comparison of the total memory accesses performed when using Alg.~\ref{alg:first_supseg} to using a simple binary search -- seeded at the middle element of the list -- for the first supersegment in a list. Fig.~\ref{fig:search-strategies} reports a comparison of the rendering frame rates obtained using different search strategies for the first supersegment in the list, binary search, linear search, and Alg.~\ref{alg:first_supseg}, on the Kingsnake (a), Rayleigh-Taylor (b) and Richtmyer-Meshkov (c) datasets. These frame rates were used to produce the speed-ups reported in Fig.~4 in the main text. The comparisons were performed on an Nvidia RTX 3090.

\begin{algorithm}
\caption{Find the first supersegment on a ray entering a list \supseglist{i} using the index of the supersegment found in the previous list
\supseglist{h} as a seed}
\KwIn{$d_{entry}$,$d_{exit}$: entry and exit intercepts of the ray in \supseglist{i}\\ 
\textit{p}: the index of the last intersected \superseg{}{} in the previous list \supseglist{h}\\ \textit{maxSupersegments} Total number of supersegments in the list \supseglist{i}}
\KwOut{index \textit{j} of the first \superseg{}{} intersected in \supseglist{i}, -1 if no such \superseg{}{}}

\begin{algorithmic}[1]
\STATE \tcc{$BinS(d,start,stop)$ is a binary search function returning an index \textit{j} such that \back{i}{j-1} $<$ $d$ $<$ \back{i}{j}, limited to $ start \leq j \leq stop$}

\STATE bin\_search\_end = -1
\STATE bin\_search\_start = -1
\IF {\textit{p} is negative}
  \STATE bin\_search\_end = maxSupersegments - 1
  \STATE bin\_search\_start = 0
\ELSE
  \STATE $b_1$ = \textbf{if}(\superseg{i}{p} exists) \back{i}{p} \textbf{else} $\infty$
  \STATE $b_0$ = \textbf{if}(\superseg{i}{p-1} exists) \back{i}{p-1} \textbf{else} $-\infty$
  \STATE interval = $b_1$ $\geq$ $d_{entry}$ + $b_0$ $\geq$ $d_{entry}$
  \IF {interval is 0}
    \STATE bin\_search\_end = maxSupersegments - 1
    \STATE bin\_search\_start = p + 1
  \ELSIF {interval is 2}
	\STATE bin\_search\_end = p - 1
	\STATE bin\_search\_start = 0
  \ELSE
    \IF {$b_1 < \infty$}
      \STATE supseg\_found = true
      \STATE \textit{index} = \textit{p}
      \STATE depthEnd = $b_1$;
    \ELSE
      \STATE bin\_search\_end = p - 1
      \STATE bin\_search\_start = 0
    \ENDIF
  \ENDIF
\ENDIF

\IF {bin\_search\_end != -1}
  \State $index = BinS(d_{entry}, bin\_search\_start, bin\_search\_end)$
\ENDIF
\State depthStart = \front{i}{index}

\IF {depthStart $> d_{exit}$}
    \State supseg\_found = false;
    \STATE $index = -1$
\ENDIF

\end{algorithmic}
\label{alg:first_supseg}
\end{algorithm}

\section{Dynamic Subsampling comparisons}
\begin{figure*}[]
    \centering
    \begin{subfigure}[t]{0.5\textwidth}
        \centering
        \includegraphics[height=2in]{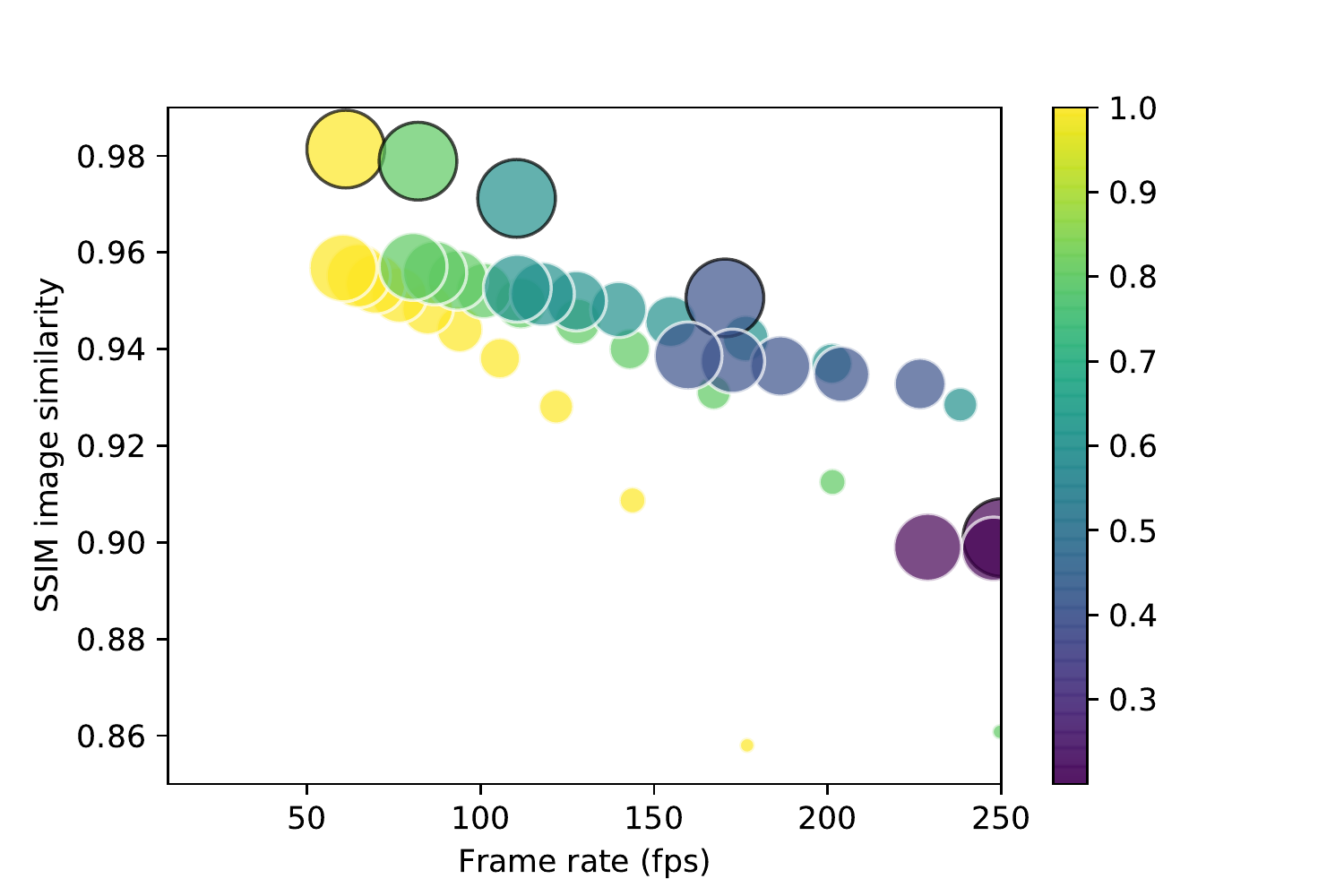}
        \caption{Kingsnake rendered at 15\textdegree~from \vieworig}
    \end{subfigure}%
    ~ 
    \begin{subfigure}[t]{0.5\textwidth}
        \centering
        \includegraphics[height=2in]{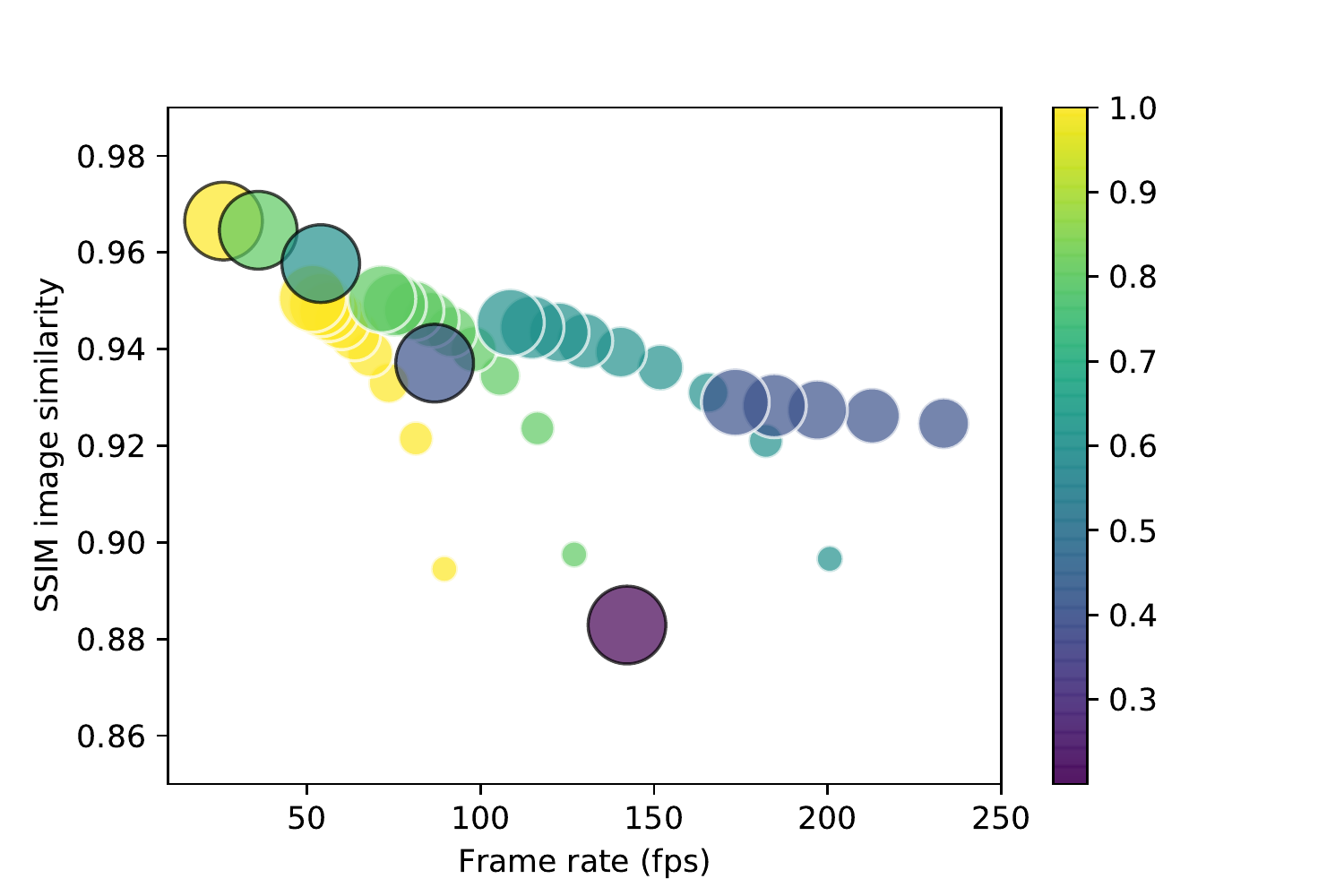}
        \caption{Kingsnake rendered at 30\textdegree~from \vieworig}
        
    \end{subfigure}
    ~
     \begin{subfigure}[t]{0.5\textwidth}
        \centering
        \includegraphics[height=2in]{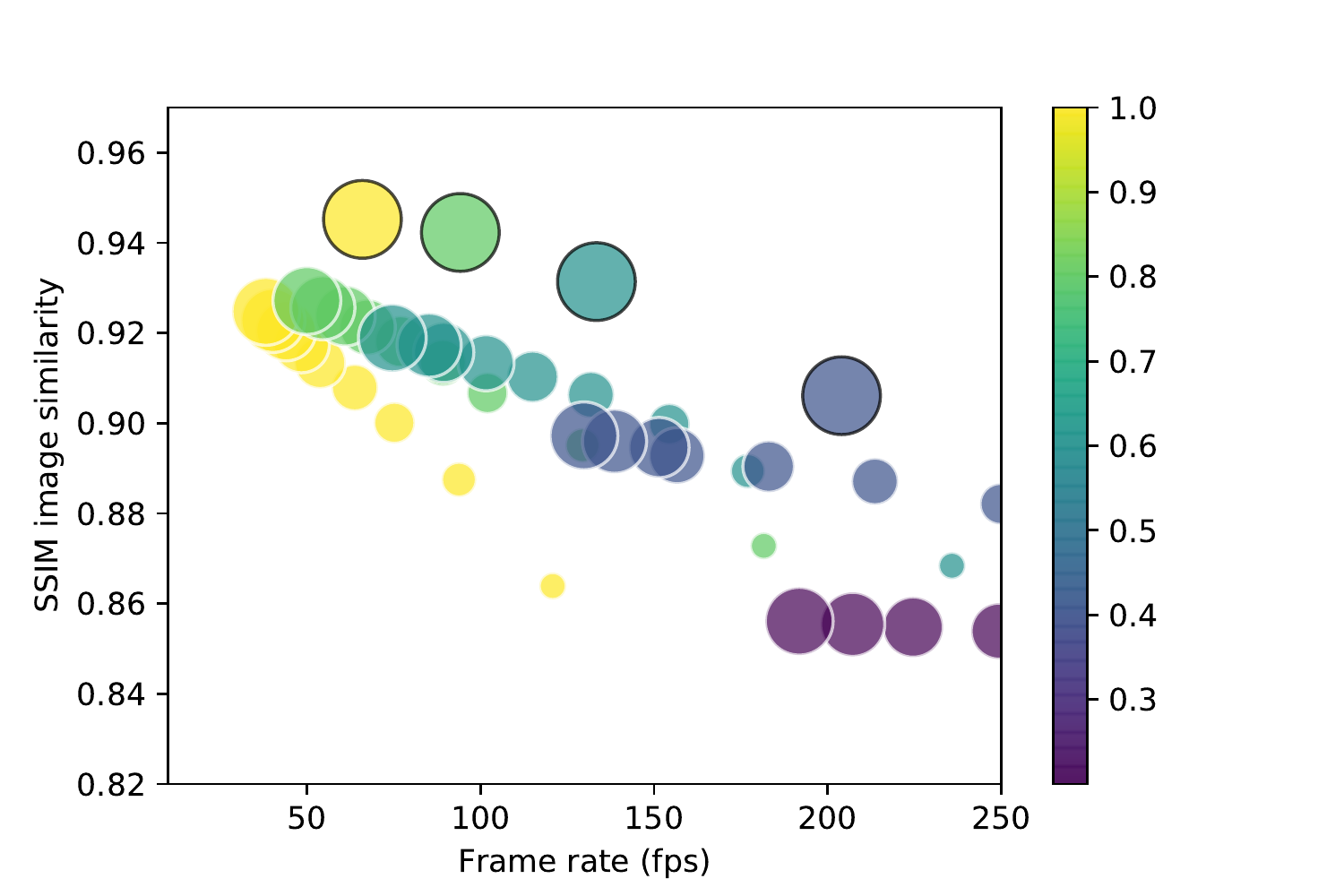}
        \caption{Richtmyer-Meshkov rendered at 15\textdegree~from \vieworig}
        
    \end{subfigure}%
    ~ 
    \begin{subfigure}[t]{0.5\textwidth}
        \centering
        \includegraphics[height=2in]{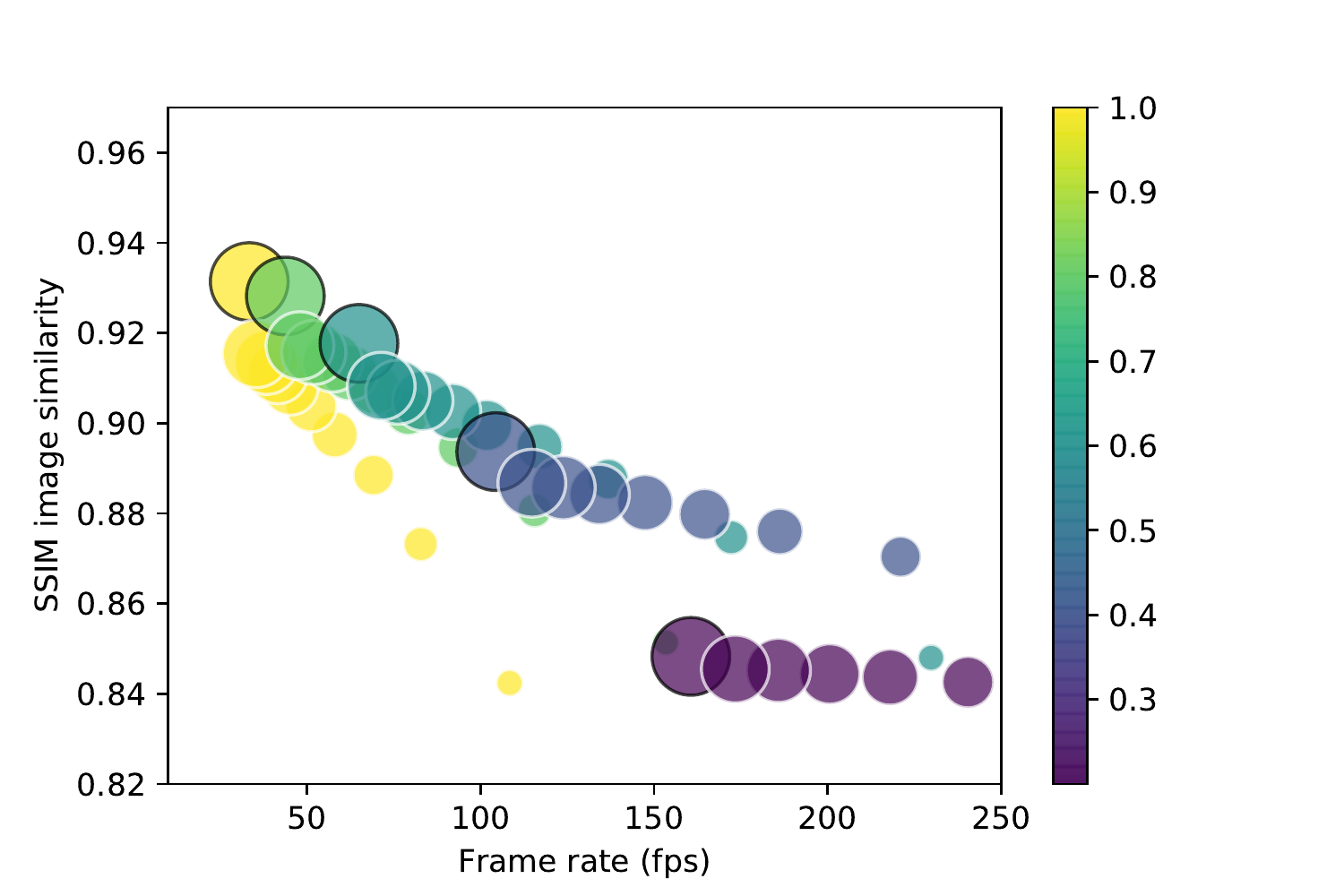}
        \caption{Richtmyer-Meshkov rendered at 40\textdegree~from \vieworig}
        
    \end{subfigure}
    
    ~
     \begin{subfigure}[t]{0.5\textwidth}
        \centering
        \includegraphics[height=2in]{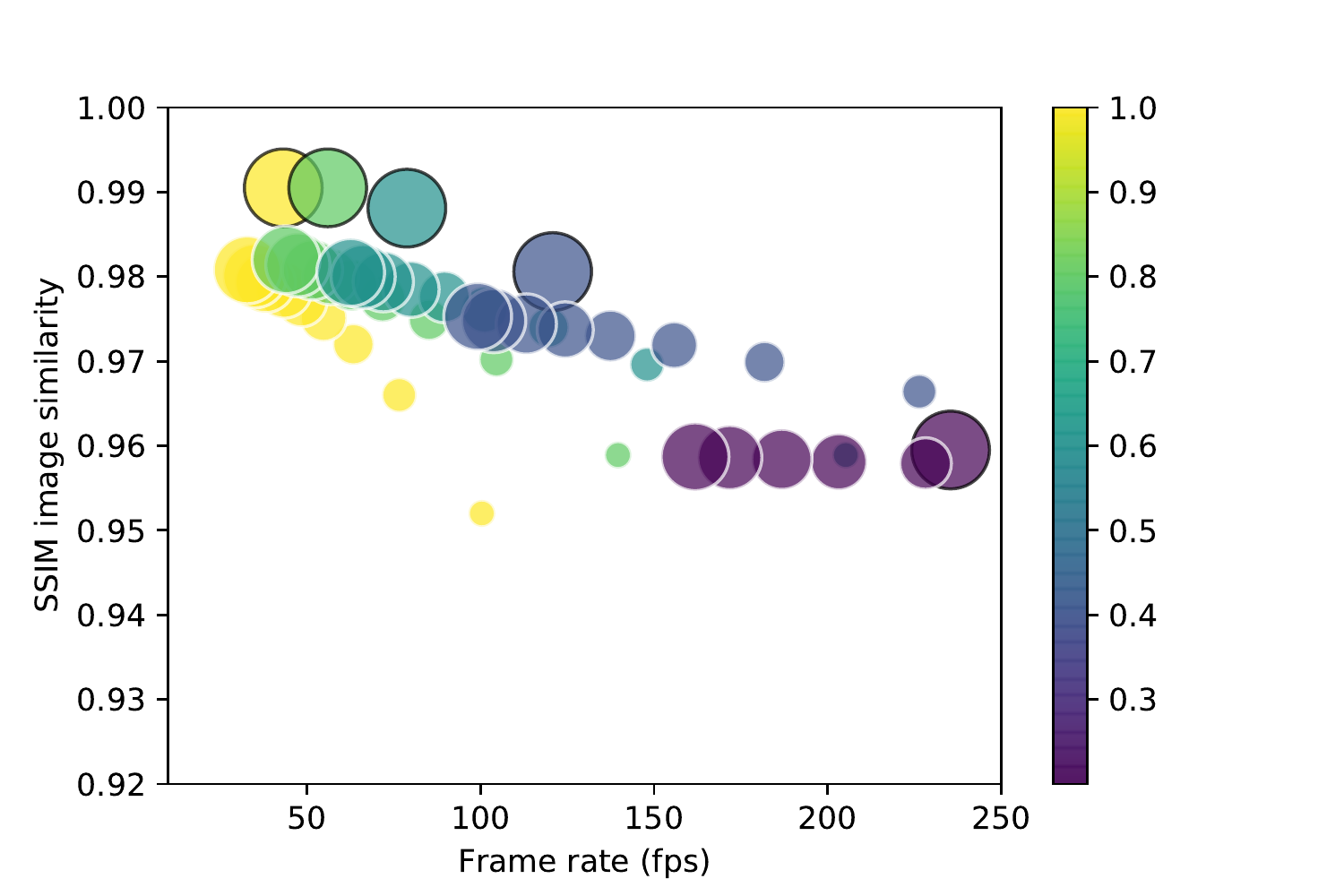}
        \caption{Beechnut rendered at 15\textdegree~from \vieworig}
        
    \end{subfigure}%
    ~ 
    \begin{subfigure}[t]{0.5\textwidth}
        \centering
        \includegraphics[height=2in]{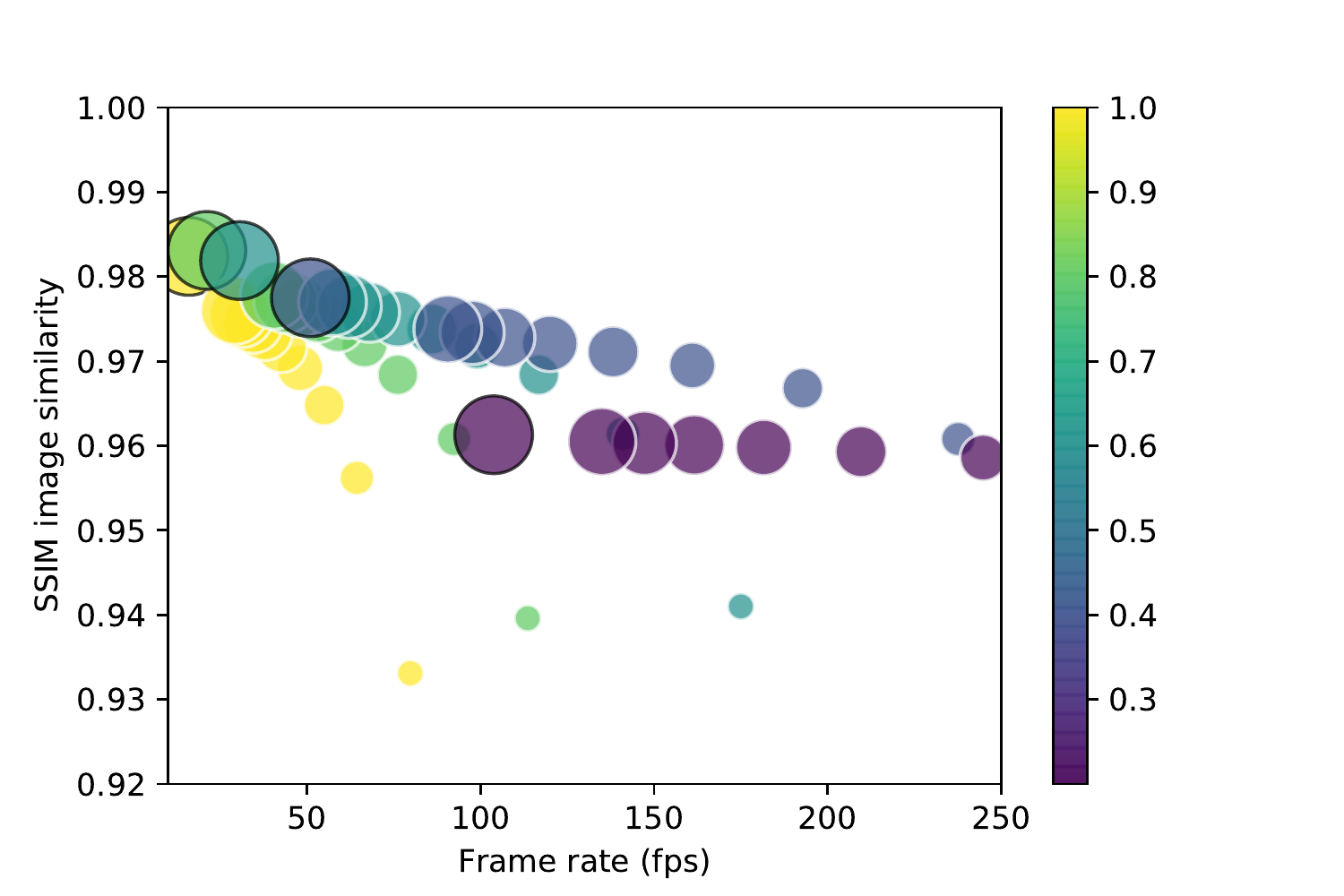}
        \caption{Beechnut rendered at 30\textdegree~from \vieworig}
        
    \end{subfigure}
    
    \caption{Analyzing the performance and quality of downsampling the rendering of the VDI along image dimensions (\downimage) and along the ray (\downray), on an RTX 3090. Color is used to represent \downimage~and smaller sizes indicate smaller \downray. Circles with black outlines indicate full-resolution rendering along the ray. Display resolution is always \fhdres, images generated with \downimage \textless 1.0 are upsampled using bilinear interpolation for display.}
    
    \label{fig:downsampling}
\end{figure*}

\begin{figure*}[t]
    \centering
    \begin{subfigure}[t]{0.5\textwidth}
        \centering
        \includegraphics[height=2in]{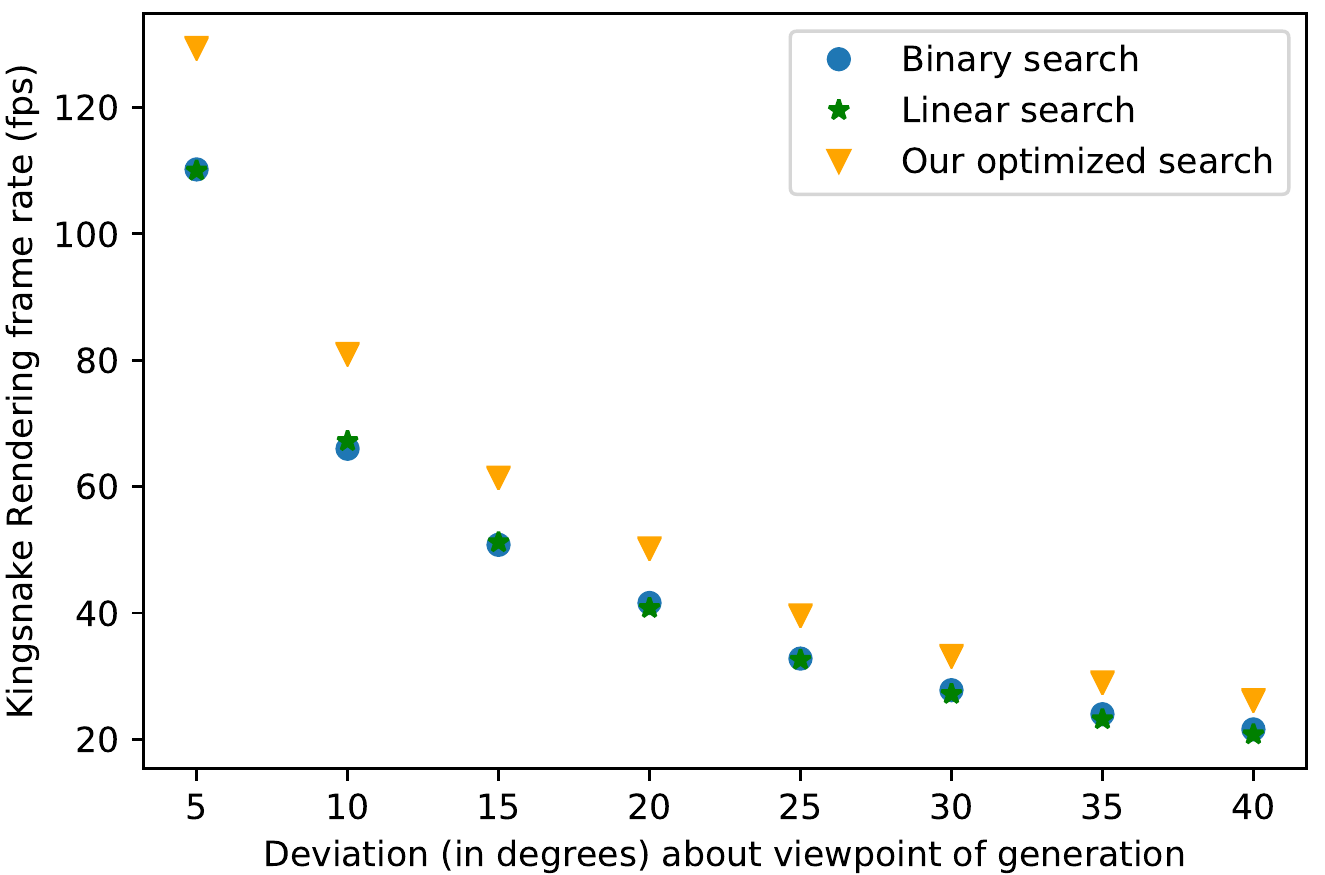}
        \caption{Kingsnake dataset.}
    \end{subfigure}%
    ~ 
    \begin{subfigure}[t]{0.5\textwidth}
        \centering
        \includegraphics[height=2in]{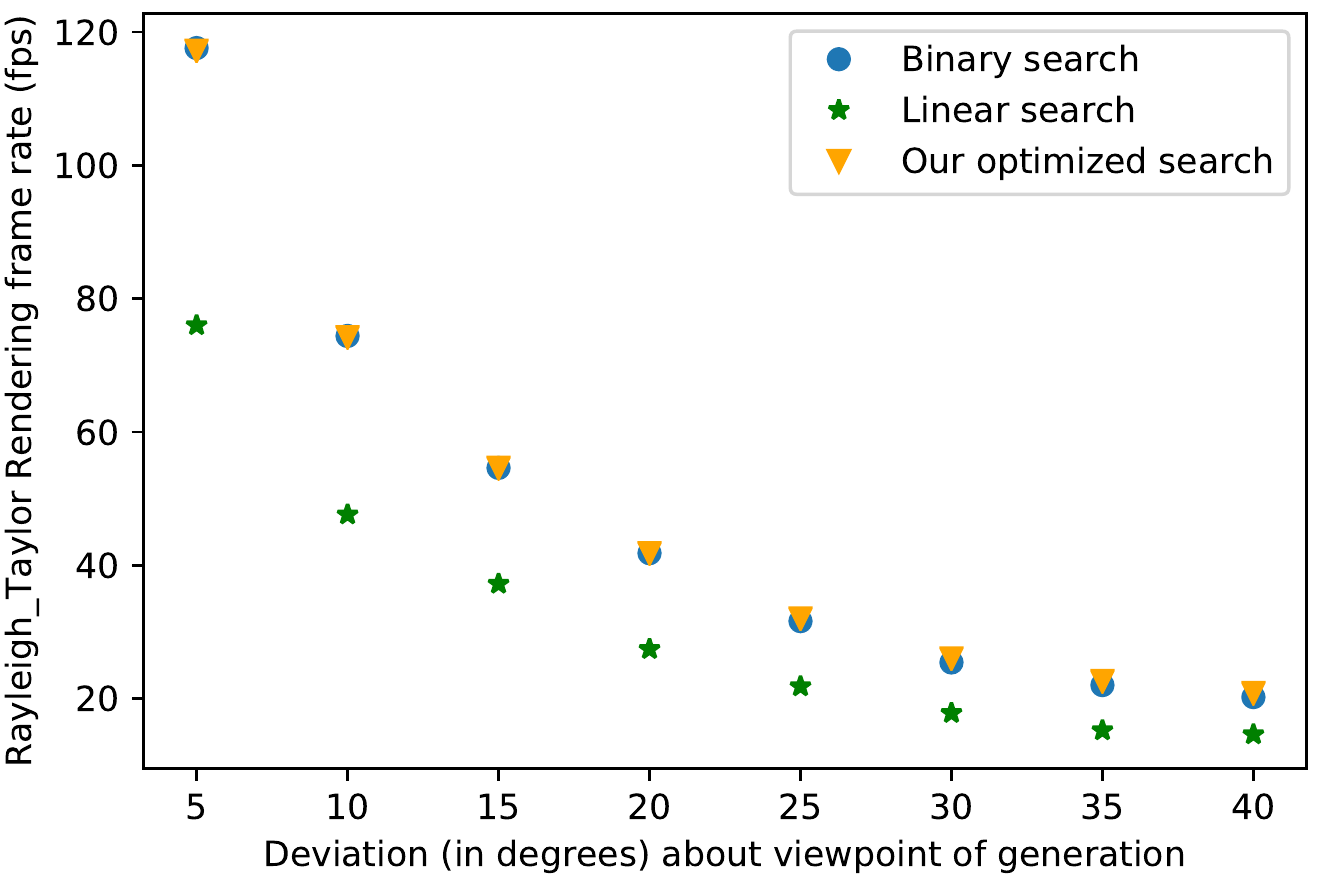}
        \caption{Rayleigh-Taylor dataset.}
        
    \end{subfigure}
    ~
     \begin{subfigure}[t]{0.5\textwidth}
        \centering
        \includegraphics[height=2in]{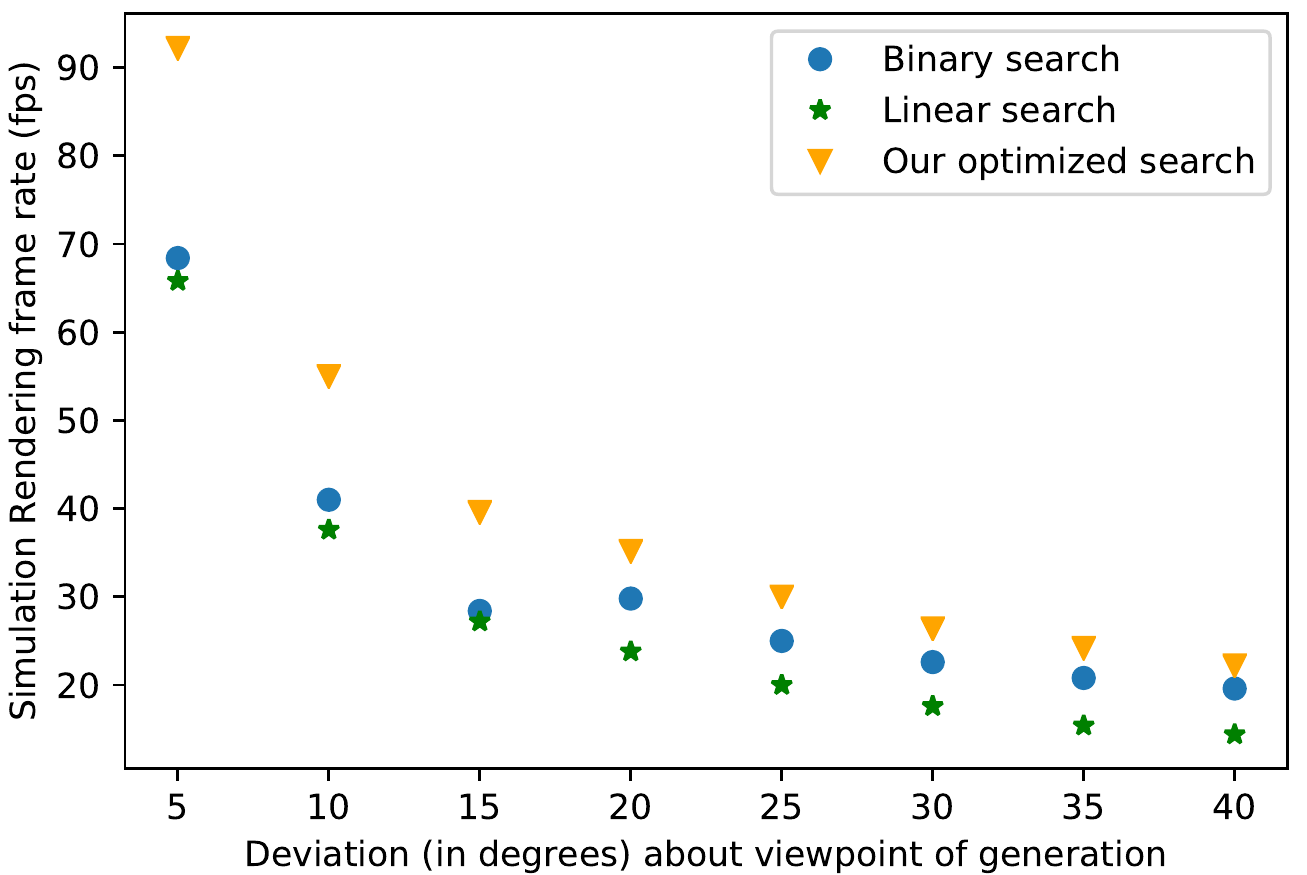}
        \caption{Richtmyer-Meshkov dataset.}
        
    \end{subfigure}%

    \caption{Comparing rendering frame rates for our optimized \superseg{}{} search (Alg.~\ref{alg:first_supseg}) against binary search and linear search strategies (symbols) for the first \superseg{}{} in a \supseglist{}. Viewport resolution=\fhdres, \numSupsegs=20.}

    \label{fig:search-strategies}
\end{figure*}

Continuing the analysis presented in Section 4.4 in the main text, Fig.~\ref{fig:downsampling} provides additional scatter plots evaluating combinations of \downimage~and \downray~for preview rendering, for multiple datasets at different view configurations. We thank the Computer-Assisted Paleoanthropology group and the Visualization and MultiMedia Lab at University of Zurich (UZH) for the acquisition of the Beechnut dataset included in the analysis here.

\section{VDI Rendering Comparison with Direct Volume Rendering}

\begin{figure}
 \centering
 \includegraphics[width=\columnwidth]{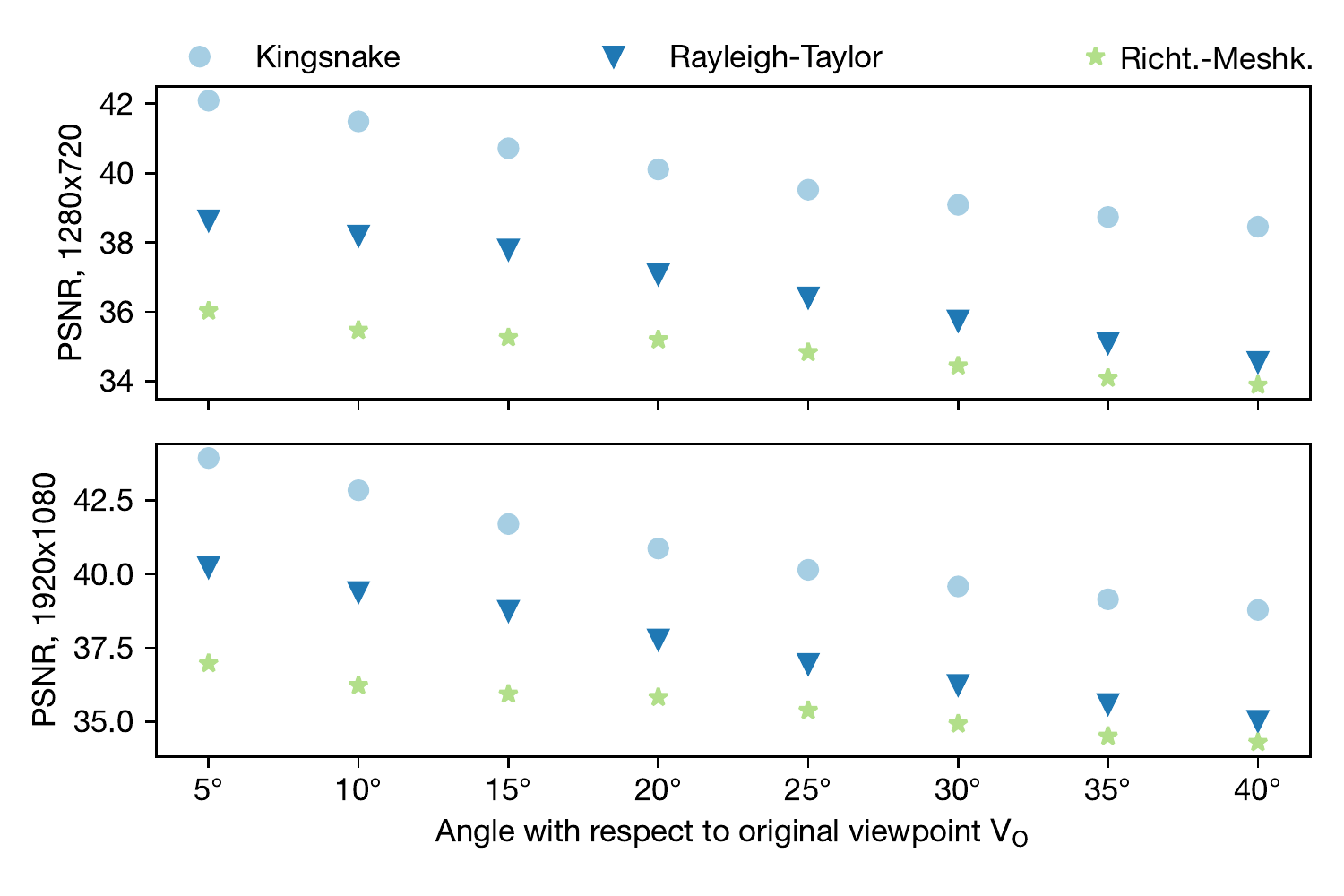}
 \caption{PSNR image similarity between VDI rendering and DVR from the same viewpoint for \shdres~(top) and \fhdres~(bottom) with \numSupsegs=20 for different datasets (symbols).}
 \label{fig:quality-psnr}
\end{figure}

Figure~\ref{fig:quality-psnr} provides a comparison of the quality of the images generated by VDI rendering with those from DVR, in terms of the PSNR (Peak Signal-to-Noise-Ratio) metric. This supplements the analysis using the SSIM metric presented in Fig.~9 in the main text. All SSIM and PSNR measurements, here and in the main text, were performed using the \texttt{scikit-image} Python package.

